\documentclass[twocolumn]{raa}

\usepackage{graphicx,times}
\usepackage{natbib}
\usepackage{amssymb,amsmath}
\bibpunct{(}{)}{;}{a}{}{,}

\usepackage[pagebackref=true]{hyperref}

\begin{document}

   \title{Research of signs of star formation in three S0 galaxies -- members of isolated pairs}

 \volnopage{ {\bf 20XX} Vol.\ {\bf X} No. {\bf XX}, 000--000}
   \setcounter{page}{1}

   \author{I.~S.~Proshina\inst{1,*}\footnotetext{$*$Corresponding Author}, O.~K.~Sil'chenko\inst{1}, A.~Yu.~Kniazev\inst{1,2,3}, D.~V.~Oparin
      \inst{4}
   }
%% Here is an example of three authors come from different institutes.
%% For single author or all the authors from an institute, use "\inst{}" only

   \institute{Sternberg Astronomical Institute of the Lomonosov Moscow State University,
University av. 13, Moscow, 119234, Russia; {\it ipro@sai.msu.ru}\\
%% Please give the E-mail address of the author, to whom future correspondence and
%% offprint requests will be sent.
        \and
             South African Astronomical Observatory, P.O. Box 9, Observatory, Cape Town 7935, South Africa\\
	\and
     	     Southern African Large Telescope, P.O. Box 9, Observatory, Cape Town 7935, South Africa\\
        \and
            Special Astrophysical Observatory of the Russian Academy of Sciences, Nizhnij Arkhyz, 369167, Russia\\
\vs \no
   {\small Received 2026 April 03; accepted 2026 October 15}
}

\abstract{We obtained and analyzed long-slit spectra of three lenticular galaxies with optical rings: PGC~48114, ESO~234-011, and 
ESO~444-076. Ionized gas was detected in all of them. In PGC~48114, the excitation mechanism of the ionized gas differs along the ring, 
despite the consistent kinematics of the ionized gas and stars, as we showed earlier. Narrow-band images of PGC~48114 in the H$\alpha$ 
and [NII]$\lambda$6583 emission lines revealed a star formation arc in the northwestern part of the ring. In ESO~234-011 and ESO~444-076, 
the ionized gas along the long slit is shock-excited and have kinematic discrepancy with the stars. 
This implies accretion of matter at an angle to the main disk of the S0 galaxies, which likely inhibits ongoing 
star formation. The companions of these S0 galaxies are late-type galaxies with extensive ongoing star formation. 
These companions may be the sources of the gas, or each pair may be feeding from a common gas reservoir.
}

   \authorrunning{I. S. Proshina et al. }            %author_head in even pages
   \titlerunning{Signs of star formation in S0 from isolated pairs}  %title_head in odd pages
   \maketitle

%________________________________________________ sections below
%
\section{Introduction}           %% first-level sections will be auto-capitalized
\label{sect:intro}

The current paradigm has been established that the evolution of disk galaxies is governed by cold gas accretion from outside 
(e.g. \citealt{Bouche+etal+2010}). The gas accretion may be permanent and laminar as in spiral galaxies, providing their smooth 
star formation history and angular momentum acquisition (\citealt{Kauffmann+etal+2010, Bouche+etal+2010, Lilly+etal+2013, 
Peng+Renzini+2020}), or intermittent -- as perhaps in lenticular galaxies (\citealt{Tacconi+etal+2020, Salim+etal+2012}). 
The difference in the gas accretion regime -- and/or in the gas inflow geometry -- may result in various final morphology of a galaxy 
having accepted the gas (\citealt{Peng+Renzini+2020}). To detalize scenarios of disk galaxy evolution, it is then important in particular 
to study lenticular galaxies with some signs of recent star formation (SF) or with evident sources of outer gas accretion. 
Lenticular galaxies with outer optical rings, seen sometimes in UV, are the most promising targets. 
There are known three types of proposed mechanisms that can probably all work and may be responsible for the origin of rings in 
galaxies. Initially, the idea that rings form at radii corresponding to the resonances of solid-body rotating bars (triaxial structures 
in the centers of galaxies) was the most popular; within this model stellar rings were formed due to the accumulation of gas at the 
resonance radii and subsequent star formation. This idea was checked by a number of successful models, e.g. \cite{Schwarz+1981}, 
and was further supported by observational statistics on the relative sizes of rings and bars (\citealt{Athanassoula+etal+1982, Buta+1986}). 
However, in parallel, the scenario of forming ring structures in the disks of galaxies was also developing related to the vertical 
passage of a satellite through the disk -- the so-called impact scenarios (\citealt{Theys+Spiegel+1977, Few+Madore+1986}). 
Resonant and impact scenarios are united by the fact that initially the galaxy's own gas gathers into the rings within the disks, and only 
then, in the process of stimulated star formation, stellar rings are formed from this gas. During the development of observational studies 
of lenticular galaxies, it became clear that in a significant majority of S0s, the gas component does not rotate together with the stellar 
one (\citealt{Kuijken+etal+1996, Davis+etal+2011}); especially in a sparse environment, about half of the gaseous disks of S0s have spins 
turned relative to the spins of the stellar disks by more than 90 degrees (so called counter-rotating gas disks) (\citealt{Katkov+etal+2015}). 
This means that in half of the cases, the ring gaseous structures in S0 galaxies cannot have either a resonant or impact origin, but can 
only be formed by accretion of gas from outside, either from another galaxy or from cosmological filaments of the large-scale structure of 
the Universe in general. The existence of accretion outer rings in galaxies was first suggested by \cite{Buta+Combes+1996}, and recently 
we have found many examples of such structures associated most often with the acquisition of matter from a small gas-rich satellite 
(\citealt{Proshina+etal+2020, Sil'chenko+Moiseev+2020}). In these cases, the gaseous rings of galaxies inherit the orbital moment of the 
satellite. They lie in the plane of the stellar disk if and only if the satellite initially has orbited coplanar to the host disk.
By establishing a relation between gas accretion conditions and the inrinsic properties of a galaxy structure and stellar populations, 
we can reveal evolutionary consequences of various outer accretion regimes.

We continue our investigation of lenticular galaxies with (outer) optical rings, also visible in UV (\citealt{Proshina+etal+2019, 
Proshina+etal+2020, Proshina+etal+2024, Sil'chenko+etal+2019}), to study the properties of ionized gas and its origin in them, and 
in this paper we consider three S0 galaxies which are parts of isolated pairs where the companion is a late-type galaxy (LTG). 
Traditionally, LTGs exhibit extensive ongoing SF, while S0 galaxies typically exhibit very modest or 
no current SF (\citealt{Chen+etal+2024, Chen+etal+2025}). However, the presence of UV-bright stellar structures beyond 
the main old disks of S0-galaxies indicates recent SF in their outermost regions. 
In addition, \cite{Tous+etal+2023}, based on the integral field spectroscopic maps (e.g. in the equivalent width of the H$\alpha$ 
emission line) from the MaNGA survey, revealed the existence of inner rings demonstrating ongoing SF in a number of S0s.

The cause of SF quenching remains unresolved and debated (\citealt{Boselli+etal+2022, Peng+Renzini+2020}), especially in the field S0 
galaxies (which represent a majority of S0s in the nearby Universe). 
The most popular mechanism for the formation of S0s in high-density environments (galaxy groups and clusters), leading to the appearance 
of so-called “faded spirals”, -- ram-pressure gas removing -- is inappropriate for galaxies in rare environments, such as isolated pairs, 
especially if the only companion is a LTG with active SF. 
Moreover, the field S0s contain often a significant amount of cold gas (\citealt{Welch+etal+2010}), and one must explain why this gas 
does not fuel SF in the disks. There exist several possible explanations. The group environment may produce tidal 
disturbance which turbulizes the gas, making the gaseous disk thicker and diminishing its volume density, so suppresing SF 
(\citealt{Bekki+Couch+2011, Deeley+etal+2020}). The similar effect may be due to minor merging 
(\citealt{Eliche-Moral+etal+2018, Fraser-McKelvie+etal+2018}). We (\citealt{Sil'chenko+etal+2019}) noted that a direction of gas inflow 
inclined with respect to the main galactic disk is unfavorable for subsequent SF because the gas flow intersecting the 
stellar disk potential well would be shocked and since heated; and just in S0s galaxies the gas kinematics is often decoupled from the 
stellar one betraying gas inflow that does not lie in the disk plane (\citealt{Davis+etal+2011, Ristea+etal+2022}). Another interesting 
scenario for S0 gas accretion differing from accretion onto spirals was proposed by \cite{Peng+Renzini+2020}. If the gas inflow has a high 
orbital momentum, the accreted gas would be concentrated at outer radii, beyond the main disk borders; and then the gravitation of the 
stellar disk is useless to compress the gas and to stimulate the SF. Indeed, the gaseous disks in massive S0s are especially 
extended (\citealt{Oosterloo+etal+2007}); and we have recently identified an isolated S0, NGC 6798, with a massive extended \ion{H}{I} 
disk, where the SF is suppressed due to increased angular momentum of the gaseous disk (\citealt{Sil'chenko+etal+2025}).

Studying the state of ionized gas and modern SF in S0 galaxies, located close to a late-type satellite and isolated from 
other visible objects, may be promising as concerning identification of outer gas source and detailing accretion conditions.

\begin{figure}
   \centering
   \includegraphics[width=12.0cm, angle=0]{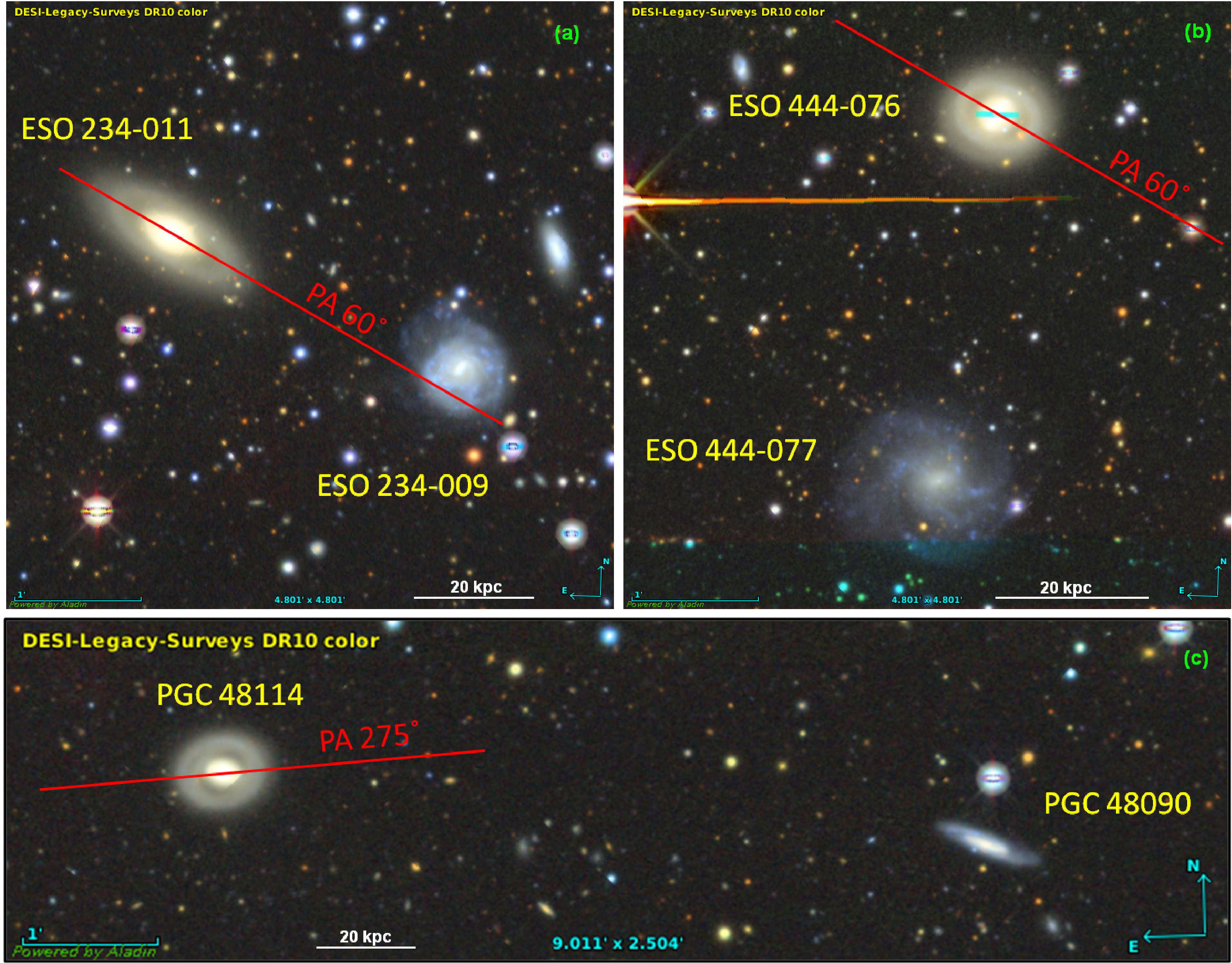}
   \caption{Optical composite-color images of isolated galaxy pairs from the Legacy Survey (\citealt{Dey+etal+2019}) 
with superimposed long slit positions.}
   \label{fig1}
   \end{figure}

\begin{figure}
   \centering
  \includegraphics[width=12cm, angle=0]{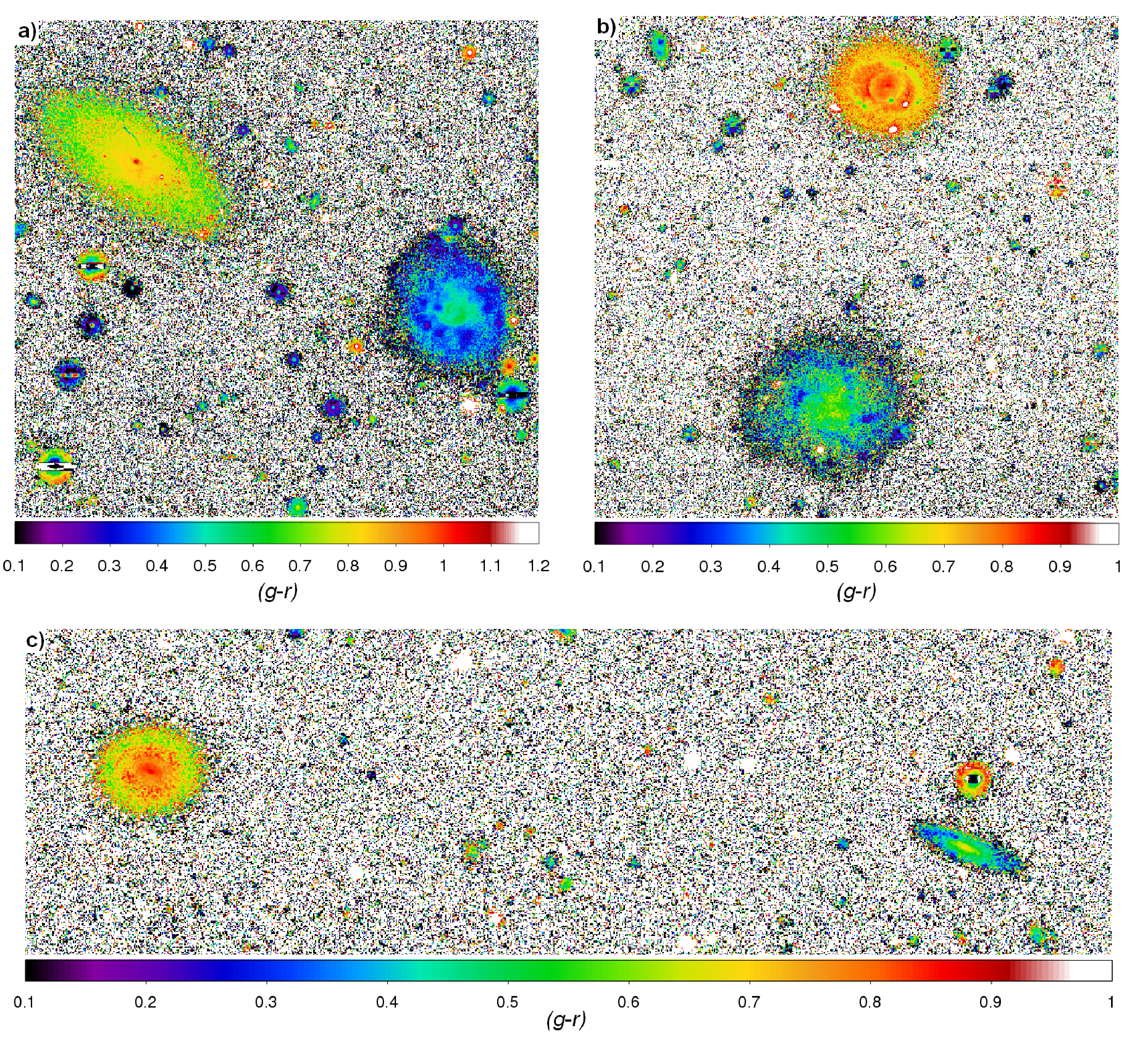}
   \caption{Color index ({\it g-r}) maps of isolated galaxy pairs: a) ESO~234-011 and ESO~234-009; 
b) ESO~444-076 and ESO~444-077; c) PGC~48114 and PGC~48090.}
   \label{fig2}
   \end{figure}

The main characteristics of the three S0 galaxies under consideration are given in Table~\ref{tab1}, and the characteristics of their 
companions are given in Table~\ref{tab2}. Figure~\ref{fig1} shows optical images of the galaxy pairs taken from the Legacy Survey 
(\citealt{Dey+etal+2019}) with the positions of long slits marked on them, and Figure~\ref{fig2} shows their color index ({\it g-r}) maps.
As can be seen from Figure~\ref{fig2}, lenticular galaxies have a higher color index ({\it g-r}) than their late-type companion galaxies, 
implying an older population in the former and a younger population in the latter, as expected. However, both types of the galaxies 
exhibit a decrease in the color index ({\it g-r}) at the periphery, which is associated with the contribution of a younger population 
and indicates evolution ‘inside-out’. In S0 galaxies, such a population is observed in ring structures. In LTGs, the 
younger population is clustered in clumps, located along spirals. Since the galaxy PGC~48090 is highly inclined to the line of sight, 
its spirals are difficult to discern. However, it is clear that the eastern side, facing its neighbor, the S0 galaxy PGC~48114, is bluer. 
The galaxy ESO~234-009 has a spiral elongated toward its neighbor, the S0 galaxy ESO~234-011. 
The galaxy ESO~444-077 appears almost undisturbed.

\begin{table}
\bc
\begin{minipage}[]{100mm}
\caption[]{Main characteristics of the S0 galaxies from literature \label{tab1}}\end{minipage}
\setlength{\tabcolsep}{7pt}
\small
 \begin{tabular}{l c c c }
  \hline\noalign{\smallskip}
Galaxy                                       &  ESO~234-011  & ESO 444-076         & PGC 48114    \\
  \hline\noalign{\smallskip}
Galaxy type (NED$^a$)                        & (R)SA(r)0$^+$ & (R$_1$)SB(rl)0$^+$  &  S?          \\
Heliocentric velocity $v_r~^a$ , km s$^{-1}$ &  5604         & 4273                &  7077        \\
Distance$^a$ , Mpc                           &  76.4         & 56.7                &  96.6        \\
Linear scale$^a$ , kpc arcsec$^{-1}$         &  0.371        & 0.275               &  0.468       \\
$R_{25}~^b$ , arcsec                         &  43.35        & 26.75               &  19.8        \\
$R_{25}$, kpc                                &  16.1         & 7.4                 &  9.3         \\
$M_B~^c$ , mag                               &  -20.14       & -19.69              &  -19.81      \\
$M_H~^a$ , mag                               &  -23.64       & -22.70              &  -23.34      \\
lg [$M_*$, M$_{\odot}$]                     &  10.90$^d$    &  10.43$^e$          &  10.37$^f$   \\
$i_{phot}~^a$, deg                           &  64.6         & 27.3                &  16.6        \\
Companion galaxy                             & ESO~234-009   & ESO 444-077         &  PGC 48090   \\
Projected distance to the companion, kpc     &  56           & 48                  &  158.7       \\
  \noalign{\smallskip}\hline
\end{tabular}
\ec
%% place \tablecomments and \tablerefs below \end{center| and \end{center}:
%% you may leave the table-width parameter to editors or set to your actual size
\tablecomments{0.86\textwidth}{$a$ -- NASA/IPAC Extragalactic Database (NED; http://ned.ipac.caltech.edu) \\
   $b$ -- Third Reference Catalogue of Bright Galaxies (\citealt{Vaucouleurs+etal+1991}) \\
   $c$ -- Lyon-Meudon Extragalactic Database (LEDA; https://leda.univ-lyon1.fr) \\
   $d$ -- calculated using formula (1) from \cite{Cluver+etal+2014} and the {\it WISE} color index {\it (W1--W2)} from the NED database \\
   $e$ -- \cite{Leroy+etal+2019} \\
   $f$ -- \cite{Moffett+etal+2015}
}
\end{table}

\begin{table}
\bc
\begin{minipage}[]{100mm}
\caption[]{Main characteristics of the companion galaxies from literature \label{tab2}}\end{minipage}
\setlength{\tabcolsep}{7pt}
\small
 \begin{tabular}{l c c c }
  \hline\noalign{\smallskip}
Galaxy                                       & ESO~234-009         & ESO~444-077       & PGC~48090   \\
\hline
Galaxy type (LEDA$^a$)                       &  Sbc                & Sd                &  Sc         \\
Heliocentric velocity $v_r~^b$ , km s$^{-1}$ &  5715               & 3865              &  6854       \\
Distance$^b$ , Mpc                           &  77.9               & 51.1              &  93.5       \\
Linear scale$^b$ , kpc arcsec$^{-1}$         &  0.378              & 0.248             &  0.454      \\
$R_{25}~^c$ , arcsec                         &  28.65              & 33.65             &  -          \\
$R_{25}$, kpc                                &  10.83              & 8.35              &  -          \\
$M_B~^a$ , mag                               &  -19.86             & -18.97            &  -19.83     \\
$i_{phot}~^b$, deg                           &  45.6               & 35.1              &  74.7       \\
lg[$M_*$, M$_{\odot}$]                      &  9.84$^d$           & 9.33$^e$          &  9.37$^f$   \\
lg[$M_{HI}$ , M$_{\odot}$]                  &  9.64$^j$           & 9.63$^h$          &  9.39$^k$   \\

\noalign{\smallskip}\hline
\end{tabular}
\ec
%% place \tablecomments and \tablerefs below \end{center| and \end{center}:
%% you may leave the table-width parameter to editors or set to your actual size
\tablecomments{0.86\textwidth}{$a$ -- Lyon-Meudon Extragalactic Database (LEDA; https://leda.univ-lyon1.fr) \\
   $b$ -- NASA/IPAC Extragalactic Database (http://ned.ipac.caltech.edu) \\
   $c$ -- Third Reference Catalogue of Bright Galaxies (\citealt{Vaucouleurs+etal+1991}) \\
   $d$ -- calculated using formula (1) from \cite{Cluver+etal+2014} and the {\it WISE} color index {\it (W1--W2)} from the NED database \\
   $e$ -- \cite{Leroy+etal+2019} \\
   $f$ -- \cite{Moffett+etal+2015} \\
   $j$ -- The HIPASS Catalog (HICAT) $+$ WISE galaxies (\citealt{Parkash+etal+2018}) \\
   $h$ -- HIPASS Bright Galaxy Catalog (\citealt{Koribalski+etal+2004}) \\
   $k$ -- The ALFALFA Extragalactic HI Source Catalog (\citealt{Haynes+etal+2018})
}
\end{table}

\section{Observations and data reduction}
\label{sect:Obs}

The long-slit spectral observations were conducted with the Robert Stobie Spectrograph (RSS) (\citealt{Burgh+etal+2003, 
Kobulnicky+etal+2003}) at the 10-m Southern African Large Telescope (SALT) (\citealt{Buckley+etal+2006, O'Donoghue+etal+2006}).
A long slit of 1.25 arcsec width was put close to the major axis for every S0 galaxy. The volume-phase grating PG0900 was used, 
which cover the spectral range of 3760 -- 6860 \AA\ with a final reciprocal dispersion of $\sim~0.97$~\AA~pixel$^{-1}$  
and FWHM spectral resolution of 5.5~\AA . The scale along the slit is 0.502~arcsec~pixel$^{-1}$. 
The log of spectral observations with a long slit is given in Table~\ref{tab3}.

\begin{table}
\bc
\begin{minipage}[]{100mm}
\caption[]{The log of spectroscopic observations of the S0 galaxies\label{tab3}}\end{minipage}
\setlength{\tabcolsep}{5pt}
\small
 \begin{tabular}{l c c c c c c}
  \hline\noalign{\smallskip}
Galaxy        & Date       &   Exposure      & Binning      &  Slit    & PA$_{slit}$ &     Seeing     \\
              &            &  (s)            &              & (arcsec) &    (deg)    & (FWHM, arcsec) \\
\hline
PGC 48114     & 2013-02-20 & 600 $\times$ 3  & 2 $\times$ 4 &  1.25    &  275        &  3.2           \\ 
\hline
ESO~234-011   & 2019-06-10 & 1400 $\times$ 2 & 2 $\times$ 4 &  1.25    &   60        &  3.4           \\
              & 2019-07-05 & 1400 $\times$ 2 & 2 $\times$ 4 &  1.25    &   60        &  3.3           \\
\hline
ESO 444-076   & 2019-07-27 & 1200 $\times$ 2 & 2 $\times$ 4 &  1.25    &   60        &  2.8           \\
\noalign{\smallskip}\hline
\end{tabular}
\ec
\end{table}

To calibrate the wavelength scale, after each galaxy observation, the spectrum of an Ar comparison arc was exposed. 
To correct for variations in sensitivity along the dispersion, spectrophotometric standard stars were observed during twilights. 
Primary data reduction was realized by means of the standard SALT science pipeline (\citealt{Crawford+etal+2010}). 
The data were then reduced according to the RSS spectral pipeline described in \cite{Kniazev+2022}. Since the effective 
field of view is 8 arcmin along the slit, we used sky spectra obtained at the edges of the slit to subtract the background.

We also obtained the narrow-band images of the galaxy pair PGC~48114 and PGC~48090 in the emission lines H$\alpha$, 
[NII]$\lambda$6583, and in the nearby red part of the continuum using the Mapper of Narrow Galaxy Lines (MaNGaL) -- photometer 
with a 13 \AA\ -bandwidth tunable filter (\citealt{Moiseev+etal+2020}) -- on the 2.5-m telescope of the Caucasus Mountain Observatory 
of the Sternberg Astronomical Institute of Moscow State University (\citealt{Shatsky+etal+2020}). 
As a detector, we used a CCD in 2022 and a CMOS in 2024, with the similar field of view, $\sim~5$~arcmin. 
The log of photometric observations with MaNGaL is given in Table~\ref{tab4}. 
Primary data reduction and fluxes calibration were done in a routine manner according to prescriptions by \cite{Moiseev+etal+2020}.

\begin{table}
\bc
\caption[]{The log of photometric observations of the galaxies PGC~48114 and PGC~48090 with MaNGaL\label{tab4}}
\setlength{\tabcolsep}{5pt}
\small
 \begin{tabular}{c c c c c c c}
  \hline\noalign{\smallskip}
Center of the filter bandpass &   Galaxy   & Date       & Total exposure  & Binning      &  Scale             &     Seeing     \\
                              &            &            &    (s)          &              & (arcsec~px$^{-1}$) & (FWHM, arcsec) \\
\hline
                              & PGC~48114  & 2024-04-09 &      2600       & 1 $\times$ 1 &     0.26           &       1.5      \\ 
 H$\alpha$                    & PGC~48090  & 2022-04-26 &      1600       & 1 $\times$ 1 &     0.33           &       1.1      \\
                              & PGC~48090  & 2022-04-27 &      2800       & 1 $\times$ 1 &     0.33           &       1.2      \\
\hline
                              & PGC~48114  & 2024-04-09 &      5200       & 1 $\times$ 1 &     0.26           &       1.5      \\ 
~[NII]$\lambda$6583           & PGC~48090  & 2022-04-26 &      1600       & 1 $\times$ 1 &     0.33           &       1.5      \\
                              & PGC~48090  & 2022-04-27 &      2800       & 1 $\times$ 1 &     0.33           &       1.2      \\
\hline
                              & PGC~48114  & 2024-04-09 &      2800       & 1 $\times$ 1 &     0.26           &       1.5      \\
Red continuum                 & PGC~48090  & 2022-04-26 &      1200       & 1 $\times$ 1 &     0.33           &       1.1      \\
                              & PGC~48090  & 2022-04-27 &      2400       & 1 $\times$ 1 &     0.33           &       1.2      \\
\noalign{\smallskip}\hline
\end{tabular}
\ec
\end{table}

\section{Data analysis and results}

\subsection{Kinematics of stars and ionized gas}

\begin{figure}
   \centering
   \includegraphics[width=13.0cm, angle=0]{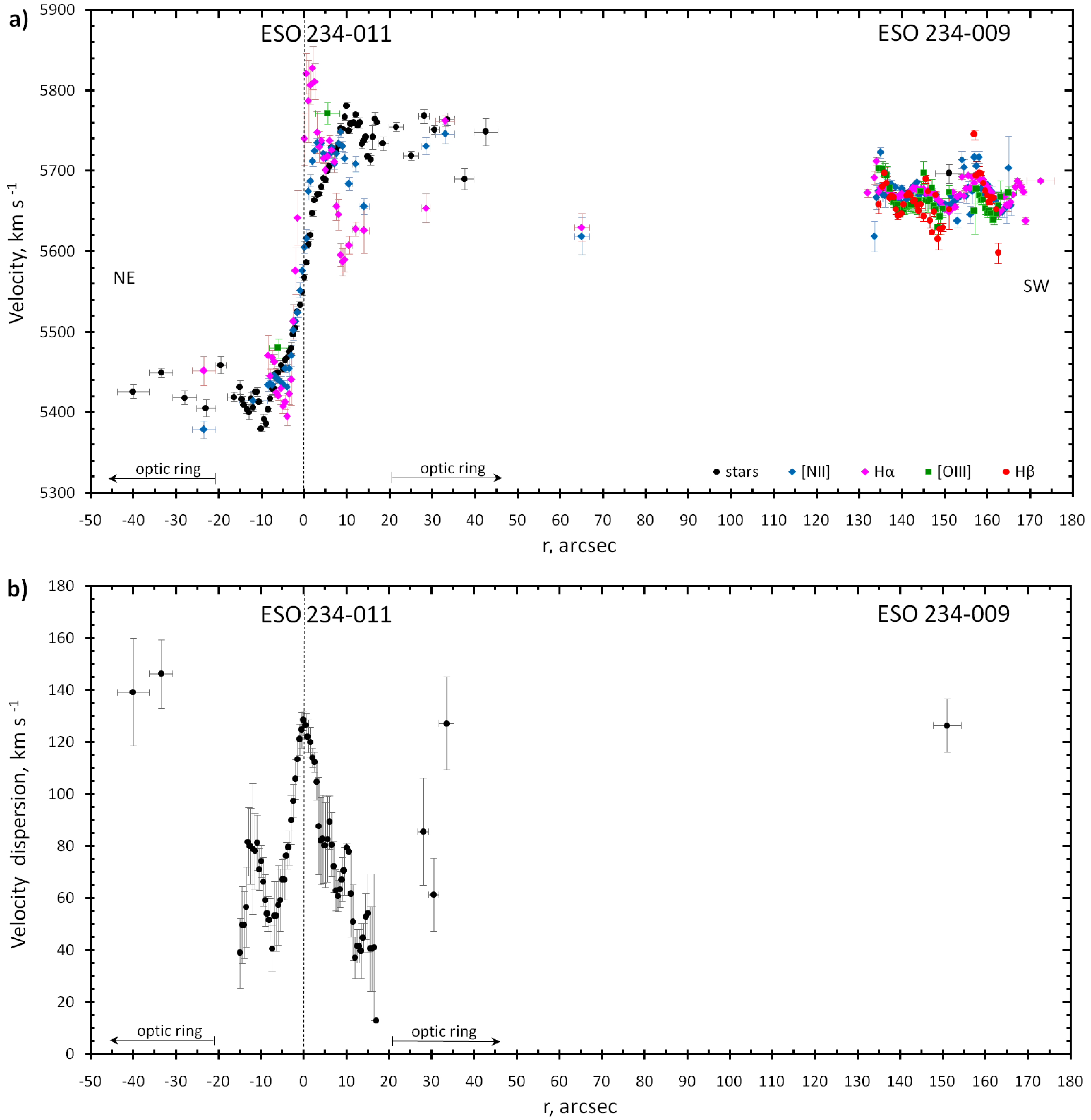}
   \caption{Variations in the line-of-sight velocities (a) and velocity dispersion (b) for the different components of ionized gas 
and stellar component along the long slit PA=$60^\circ$ in the galaxy pair ESO~234-011 and ESO~234-009.}
   \label{fig3}
   \end{figure}

\begin{figure}
   \centering
  \includegraphics[width=10cm, angle=0]{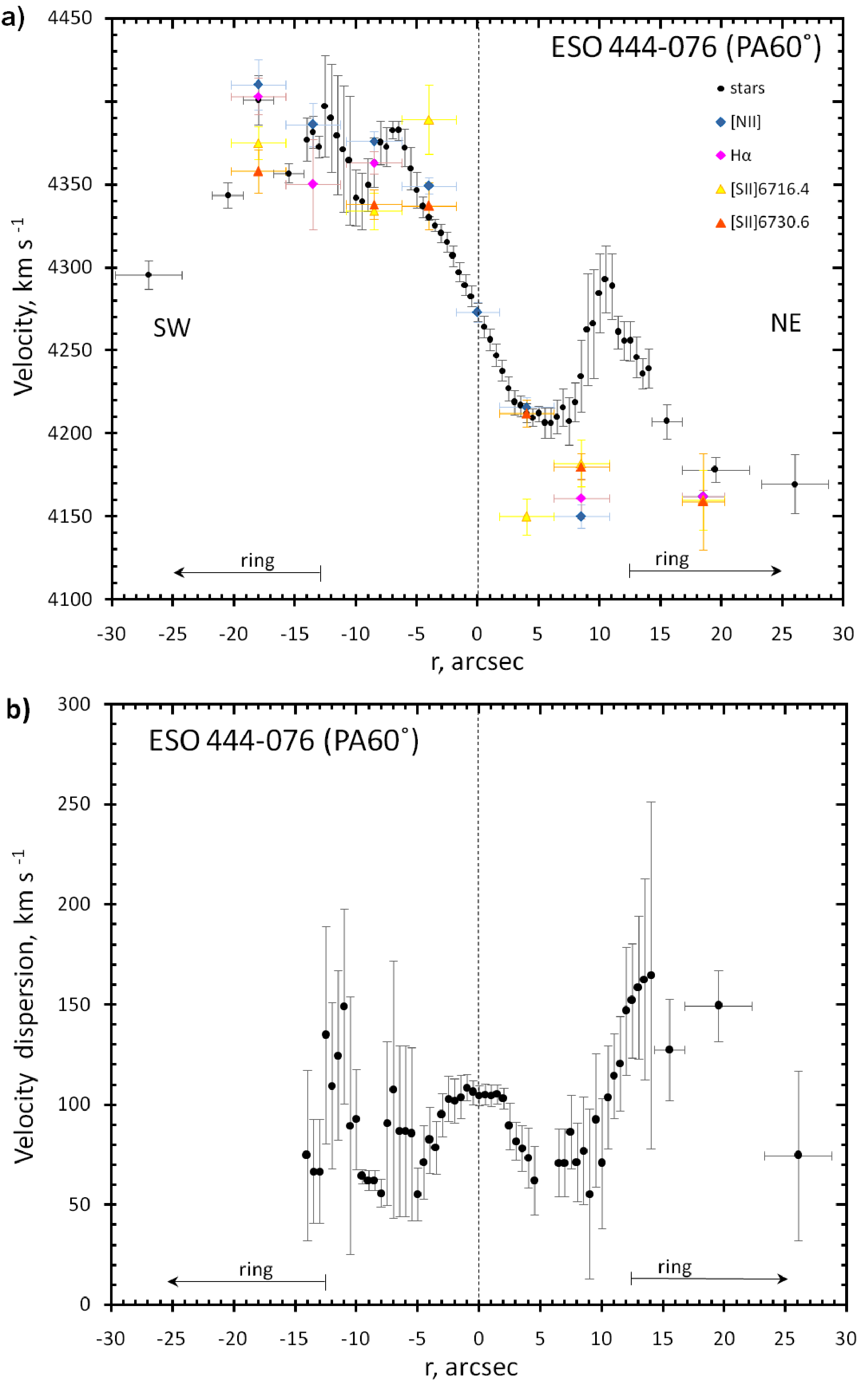}
   \caption{Variations in the line-of-sight velocities (a) and velocity dispersion (b) for the different components of ionized gas 
and stellar component along the long slit PA=$60^\circ$ in the galaxy ESO~444-076.}
   \label{fig4}
   \end{figure}

We applied Gaussian fitting to calculate the line-of-sight velocities of the ionized gas components and measure the 
equivalent widths of the emission lines. Line-of-sight velocities of the stellar component and velocity dispersion 
were derived by cross-correlating the blue-green part of the galaxy spectra with the spectrum of the bright K-type 
giant star HD58972, obtained in the same spectrograph configuration. To improve the signal-to-noise ratio with 
increasing distance from the galaxy center, the spectra were summed manually over bins. The radial summation ranges are denoted 
by horizontal bars in Figure~\ref{fig3} and Figure~\ref{fig4}, which show graphs of the ionized gas and stellar line-of-sight 
velocity curves, as well as stellar velocity dispersion, for galaxies ESO~234-011 and ESO~444-076, respectively. 
Since our long slit at PA=$60^\circ$ in the case of ESO~234-011 intersected the disk of the neighboring LTG 
ESO~234-009, located at a projected distance of about 150 arcsec from the center of ESO~234-011, the plot (Fig.~\ref{fig3}) 
also shows the measured line-of-sight velocities of ionized gas for the companion galaxy. 
Its emission lines have turned out to be much stronger than in the lenticular galaxy ESO~234-011, while the stellar 
continuum is very weak, so the line-of-sight velocity of the stellar component and the stellar velocity dispersion in 
ESO~234-009 are denoted by a single average point with error bars.

As can be seen from Figure~\ref{fig3}, the line-of-sight velocity curves of ionized gas in ESO~234-011, obtained from the 
H$\alpha$ and [NII]$\lambda$6583 emission lines, are inconsistent with each other, with the stellar component, and asymmetric 
with respect to the galactic center. 
We have to note the presence of the ‘island’ of emission lines of ionized gas between these two galaxies at a distance of 
about 65 arcsec from the center of ESO~234-011, with a line-of-sight velocity closer to the line-of-sight velocities of 
ionized gas components in the companion galaxy ESO~234-009 than in the S0 galaxy ESO 234-011. The mismatch between the 
stellar and gas kinematics in ESO~234-011, as well as the jump in the line-of-sight velocities of ionized gas on the side 
facing the neighboring galaxy, observed in the H$\alpha$ emission line at a distance of up to 10 arcsec from the center, 
implies gas accretion from the outside, most likely from the side of the neighboring late-type galaxy ESO~234-009, 
occurring at an angle to the plane of the stellar disk of the S0 galaxy ESO~234-011. The discrepancy between the 
line-of-sight velocities measured by the H$\alpha$ and [NII]$\lambda$6583 emission lines can presumably be explained by 
the fact that the H$\alpha$ emission line traces the inflowing (less enriched) gas, while the [NII]$\lambda$6583 emission 
line traces the ‘post star formation’ regions that have already virialized in the gravitational potential of the S0 galaxy 
and therefore better coincide with the line-of-sight velocity curve of its stellar component.

The line-of-sight velocity curve of stellar component in ESO~444-076, as seen in Figure~\ref{fig4}, drops sharply starting at a 
distance of 5 arcsec from the center, then rises again at 12.5 arcsec, where the ring visible in the optical and UV begins. This 
drop may be due to the presence of a nested stellar structure in the bar with retrograde orbits (\citealt{Wozniak+Pfenniger+1996}), 
or to the effect of non-circular motions of stars in the bar, which is perpendicular to the major axis of the ring, along 
which the long slit was put. At the same distances from the galactic center (beyond 5 arcsec), an increase in stellar 
velocity dispersion is observed. At the galactic center, the stellar velocity dispersion plateaus at 105\,km~s$^{-1}$. 
Overall, the kinematics of stars and ionized gas appear to be perturbed and mismatched.

Although in both S0 galaxies the ionized gas rotates in the same direction as the main stellar component, 
there are kinematic discrepancies implying an accretion origin of the gas (see, for example, \cite{Cao+etal+2022}). 
The observed increase in stellar velocity dispersion in the outer parts of the S0 galaxies ESO~234-011 and ESO~444-076 
can be supposably explained by the contribution of an accreted stellar component.

The kinematics of the ionized gas and stellar component in the S0 galaxy PGC~48114 are quite consistent and have been 
studied and described by us previously (see fig.~3 in \cite{Proshina+etal+2019}).

\subsection{Diagnostics of ionized gas in the galaxies}
\label{Diagnostics}

\subsubsection{Spectroscopy results}

\begin{figure}
   \centering
   \includegraphics[width=14.0cm, angle=0]{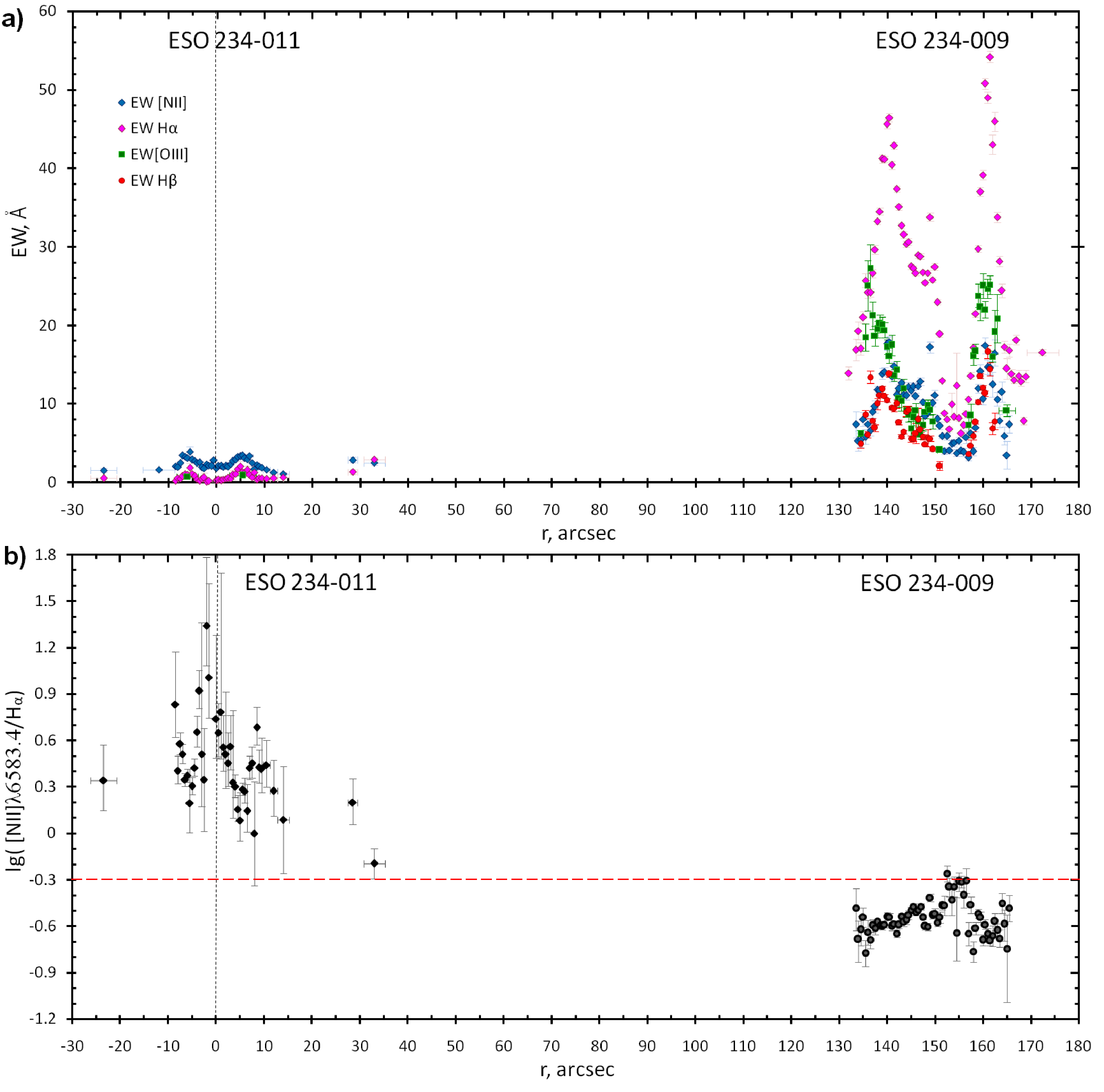}
   \caption{Variation of the equivalent widths (EW) of emission lines along the long slit ({\it top}) and 
a variation of $\lg(\mbox{[NII]6583/H}\alpha)$ ({\it bottom}) in the galaxies ESO~234-011 and ESO~234-009.}
   \label{fig5}
   \end{figure}

Ionized gas in ESO 234-011 can be seen both in the central part of the galaxy out to a radius of 15 arcsec from the 
center, and in the outer ring region (at radii of 24 -- 48 arcsec from the center), where emission lines are observed 
only in places. Figure~\ref{fig5} presents plots of the variation of the equivalent widths of emission lines along the long slit 
({\it top}) and a graph of the variation of $\lg(\mbox{[NII]6583/H}\alpha)$ ({\it bottom}) in the galaxies ESO~234-011 
and ESO~234-009.

As can be seen from Figure~\ref{fig5}, the maxima of the equivalent widths of emission lines in ESO~234-011 are located at a 
distance of 6 arcsec on either side of the center, i.e. in the same ring-like region where a plateau in the line-of-sight 
velocities measured by the [NII] emission lines is observed in the range of 3 -- 8 arcsec from the center, and where 
a weak [OIII]$\lambda$5007 emission line is detected. This region lies inside the inner ring, starting at a distance 
of 10 arcsec from the center (see below in Sec.~\ref{structure} and \cite{Grouchy+etal+2010}), i.e. the emission peaks 
are observed in front of the inner edge of the inner ring of the galaxy ESO~234-011. 
The strongest line in this galaxy is the nitrogen [NII]$\lambda$6583, and the equivalent widths of the H$\alpha$ emission 
line do not exceed 2~\AA\ , which means that in ESO~234-011 (along the position angle PA=$60^\circ$) there is no current star 
formation, according to the criterion of \cite{Lacerda+etal+2018}, and the gas there is most likely ionized by shock mechanisms 
(\citealt{Allen+etal+2008}) occurring in front of the inner edge of the inner ring of the galaxy, and/or by the light of old stars 
(\citealt{Binette+etal+1994}).

The sulfur doublet's emission line [SII]$\lambda$6716.4 turned up at the edge of the spectrograph's spectral range, 
and due to the galaxy's rotation, this line was partially truncated to the right of the galaxy's center. However, the 
average line-of-sight velocity to the left of the center for the [SII]$\lambda$6716.4 emission line is 5430\,km~s$^{-1}$ and is 
consistent with values obtained for other emission lines at same distances. Meanwhile, our measurements for the sulfur emission 
line are not shown in Figure~\ref{fig5} due to uncertain identification.

\begin{figure}
   \centering
  \includegraphics[width=10cm, angle=0]{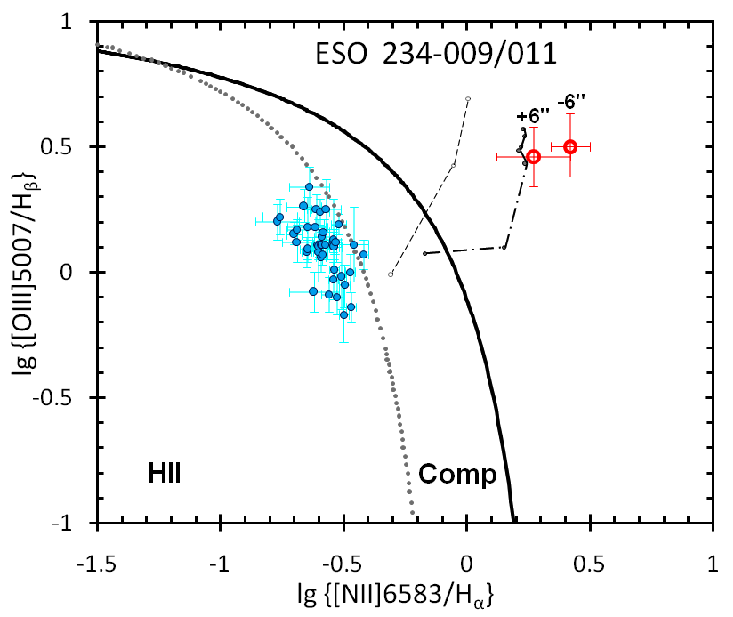}
   \caption{Comparison of the excitation mechanisms of ionized gas in the ring-like region at a radius of about 6 arcsec in the galaxy 
ESO~234-011 ({\it red symbols}) and in the disk of ESO~234-009 ({\it blue dots}) using the BPT diagram from \cite{Kewley+etal+2006}.  
The theoretical model curve and the empirical boundary curve are constructed using the equations from \cite{Kewley+etal+2006} 
and are shown as solid and dotted lines, respectively. The models of shock excitation of ionized gas with a density of 1 cm$^{-3}$ 
and solar oxygen abundance are plotted, according to calculations by \cite{Allen+etal+2008}: the dashed-dotted line corresponds 
to pure shock excitation, while the dashed line corresponds to shock excitation with a precursor.}
   \label{fig6}
   \end{figure}

In the neighboring galaxy ESO~234-009, on the contrary, along the part of the disk crossed by the long slit, the emission 
in the H$\alpha$ line dominates, and the values of the equivalent widths of all emission lines are several times and even 
tens of times higher than in the S0 galaxy ESO~234-011. The plot of $\lg(\mbox{[NII]6583/H}\alpha)$ shown in Figure~\ref{fig5} 
({\it bottom}) confirms and clearly demonstrates that SF is ongoing in the LTG ESO~234-009, since 
almost all $\lg(\mbox{[NII]6583/H}\alpha)$ values are below the criterion value of $\lg(\mbox{[NII]6583/H}\alpha)=-0.3$ 
(indicated by the red horizontal dashed line in the plot), adopted based on the BPT diagrams from \cite{Kewley+etal+2006} for 
\ion{H}{II} regions; the same plot confirms the absence of current SF in the S0 galaxy ESO~234-011, since all 
$\lg(\mbox{[NII]6583/H}\alpha)$ values exceed the criterion value of $\lg(\mbox{[NII]6583/H}\alpha)=-0.3$. These conclusions 
are also confirmed by the BPT diagram we constructed, shown in Figure~\ref{fig6}. Two red symbols indicate the state of ionized gas 
in the ring-like region of ESO~234-011 at a distance of about 6 arcsec on either side of the center (i.e., where the EWs are 
maximum, where a plateau in the line-of-sight velocities measured by the [NII]$\lambda$6583 emission line is observed, and 
where a weak [OIII]$\lambda$5007 emission line is detected). It is clear that both red symbols lie in the locus of shock-induced 
gas excitation. The blue dots mark the values for ionized gas in the late-type galaxy ESO~234-009: all of the dots fall in the 
locus of an \ion{H}{II} region.

It is worth noting that in the isolated emission ‘island’ between these two galaxies the continuum is too weak, and 
H$\alpha$ and [NII]$\lambda$6583 emission lines are weak but measurable, so $\lg(\mbox{[NII]6583/H}\alpha)=-0.16 \pm 0.20$, i.e. 
the gas there has a composite excitation mechanism.

Since the LTG ESO~234-009 is confidently undergoing star formation, this allows us to estimate the metallicity 
of the gas. Thus, the average oxygen abundance of the gas, with the standard error of the mean, 
estimated by the {\it N2} indicator, is $12+\lg$(O/H) = $8.59 \pm 0.06$ (in the calibration of \cite{Pettini+Pagel+2004}) and 
$12+\lg$(O/H) = $8.49 \pm 0.05$ (in the calibration of \cite{Marino+etal+2013}); and by the {\it O3N2} indicator: 
$12+\lg$(O/H) = $8.51 \pm 0.05$ (in the calibration of \cite{Pettini+Pagel+2004}) and $12+\lg$(O/H) = $8.39 \pm 0.03$ (in the 
calibration of \cite{Marino+etal+2013}). That is, the gas in the disk of a late-type star-forming galaxy ESO~234-009 has a metallicity 
slightly lower than the solar one, given by \cite{Asplund+etal+2009} as $12+\lg$(O/H)$_{\odot}$=$8.69 \pm 0.05$.

\begin{figure}
   \centering
  \includegraphics[width=12cm, angle=0]{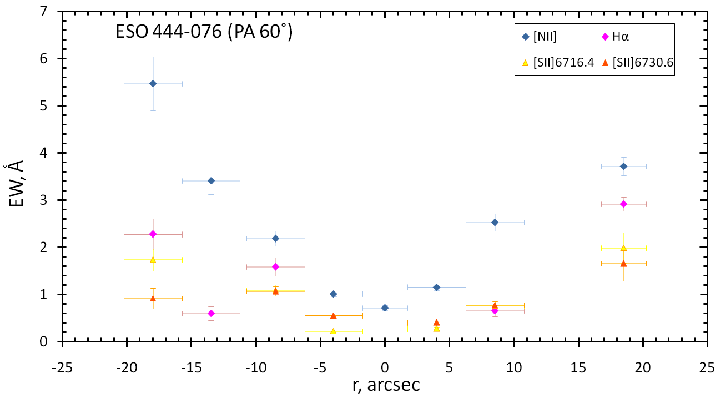}
   \caption{Variation of the equivalent widths (EW) of emission lines along the long slit in the galaxy ESO~444-076.}
   \label{fig7}
   \end{figure}

Figure~\ref{fig7} shows the variation of the equivalent widths of the emission lines in the S0 galaxy ESO~444-076, obtained along 
the slit PA=$60^\circ$. It is evident that at this slit position, the emission in the ionized gas is weak throughout the galaxy: 
it almost disappears toward the center and increases in the ring region (12.5 -- 25 arcsec from the center), with the nitrogen 
doublet emission lines [NII] dominating everywhere. The equivalent widths of the H$\alpha$ emission line are less than 3~\AA\ , 
and, therefore, along the position angle PA=$60^\circ$ in ESO~444-076 there is no current star formation according to the criterion 
of \cite{Lacerda+etal+2018}, as in case of the S0 galaxy ESO~234-011.

In the case of the galaxy ESO~444-076, both emission lines of the sulfur doublet [SII]$\lambda,\lambda$6717,6731 fell into 
the spectral range of the spectrograph, that allows us to estimate the electron density along the long slit (except for the 
center of the galaxy, where the sulfur emission lines are absent) based on the flux ratio in these lines, using the nomogram 
from \cite{Osterbrock+Ferland+2006}. Thus, at a radius of 4 arcsec from the center, the electron density $n_e$ is maximal and 
equal to 2200 -- 3000 cm$^{-3}$, and then, as we move away from the center, it decreases to a value of 550 cm$^{-3}$ at a 
distance of 8.5 arcsec from the center and then, at a distance of 18.5 arcsec, it drops to values of 0 -- 250 cm$^{-3}$.

\begin{figure}
   \centering
  \includegraphics[width=11cm, angle=0]{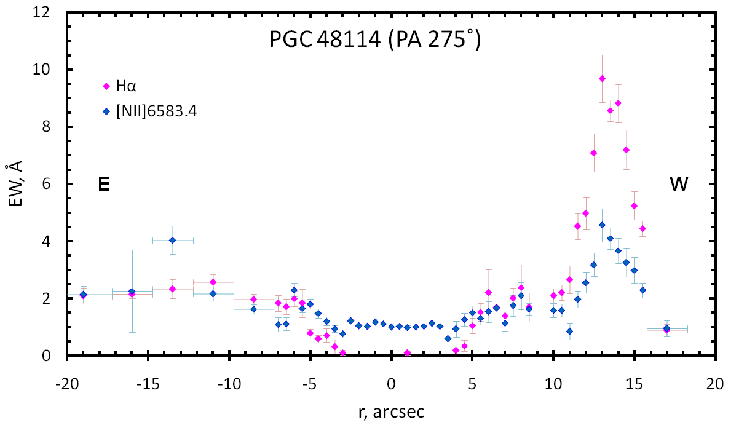}
   \caption{Variation of the equivalent widths (EW) of emission lines along the long slit PA=$275^\circ$ in the galaxy PGC~48114.}
   \label{fig8}
   \end{figure}

As can be seen from Figure~\ref{fig8}, which shows the variations in the equivalent widths of the emission lines along the long slit 
in the galaxy PGC~48114, the peak of the equivalent widths of the emission lines is located at a distance of 13 arcsec on either side 
of the center, i.e. in the region of the ring visible in the optical range, and only in the western half of the ring is current star 
formation diagnosed, since H$\alpha$ emission dominates there and $\lg(\mbox{[NII]6583/H}\alpha)<-0.3$. This prompted us to 
perform narrow-band photometry of the pair PGC~48114 and PGC~48090 to obtain maps of the distribution of ionized gas in them 
and to determine the mechanisms of its excitation over the entire area of their disks not crossed by the long slit.

The emission lines of the sulfur doublet turned up outside the spectral range of the spectrograph, and the emission lines 
[OIII]$\lambda$5007 and H$\beta$ are absent in the spectrum of PGC~48114.

\subsubsection{Narrow-band photometry results for the pair PGC~48114 and PGC~48090}

\begin{figure}
   \centering
  \includegraphics[width=14.0cm]{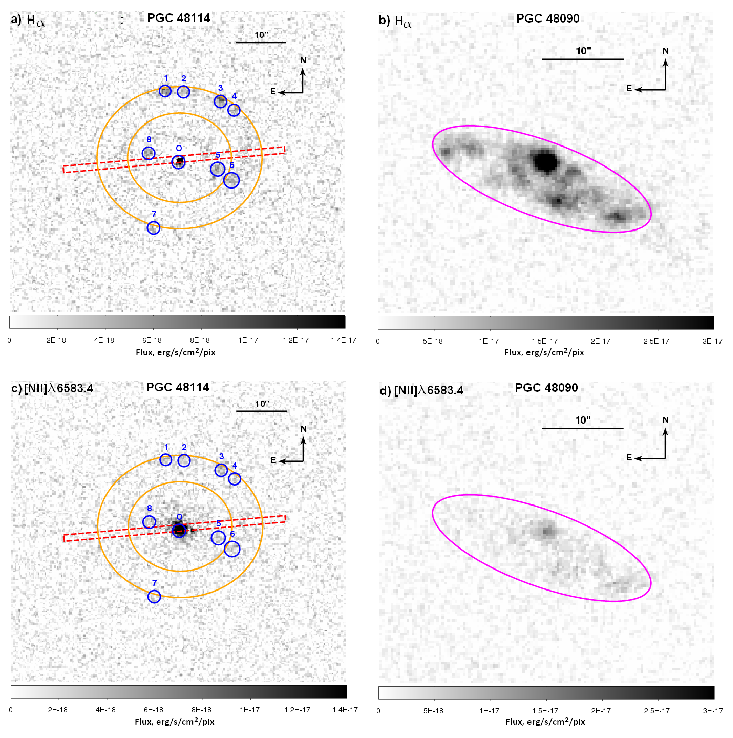}
   \caption{Intensity maps for PGC~48114 ({\it left}) and PGC~48090 ({\it right}) obtained with the MaNGaL in narrow emission lines: 
(a), (b)--H$\alpha$, (c), (d)--[NII]$\lambda$6583 with continuum subtracted. The apertures of the ring in PGC~48114 and the ellipse 
of the disk in PGC~48090, in which fluxes were measured, are superimposed. Blue circles indicate emission clumps. 
The red dotted rectangle indicates the position of the long slit during the spectroscopic observations.}
   \label{fig9}
   \end{figure}

Figures~\ref{fig9}a--d show the derived intensity maps of the emission H$\alpha$ and [NII]$\lambda$6583 lines with continuum subtracted 
for PGC~48114 and PGC~48090. Several clumps stand out in the maps for PGC~48114, four of which are located in the northwest along 
the outer edge of the stellar ring visible in optical images and indicated by the orange ring-like aperture. The bright central 
clump and three other clumps are located almost along the major axis of the central stellar body of the galaxy PGC~48114, and 
another clump is located to the south-southeast on the outer edge of the ring. 
In the late-type galaxy PGC~48090, the brightest clump is offset from the center to the northern part of the disk. 
Given their relative positions, and the fact that the system velocity of the late-type galaxy is 223 km/s lower than the system 
velocity of the S0 galaxy (which may be caused by peculiar velocities), it can be assumed that the observed shift of the brightest 
star-forming region was caused by the passage of a late-type galaxy near the S0 galaxy some time ago. 
But there are no clumps of ionized gas outside these two galaxies.

For each selected clump in PGC~48114, we measured the H$\alpha$ emission line flux, the neighboring continuum flux, and the 
[NII]$\lambda$6583 emission line flux. We applied a flux correction that takes into account the partial overlap of the wings 
of the transmission profile of the filter centered on these emission lines, as described by \cite{Moiseev+etal+2020}, and 
calculated the mean equivalent widths (EW) of the H$\alpha$ emission line (multiplying by the filter bandwidth -- 13 \AA\ -- and 
dividing by the fluxes in the neighboring continuum) and values of the logarithm $\lg(\mbox{[NII]6583/H}\alpha)$. For those clumps 
in which we diagnosed current star formation, because of the $\lg(\mbox{[NII]6583/H}\alpha)<-0.3$, we estimated the oxygen abundance 
in the gas using the {\it N2} indicator in two calibrations: \cite{Pettini+Pagel+2004} and \cite{Marino+etal+2013}. 
We performed the same procedures for the entire disk of the late-type galaxy PGC~48090, outlined by the ellipse in 
Figures~\ref{fig9}b,d. The results are presented in Table~\ref{tab5}.

\begin{table}
\bc
\caption[]{Characteristics of emission regions in galaxies PGC~48114 and PGC~48090\label{tab5}}
\setlength{\tabcolsep}{2pt}
\small
 \begin{tabular}{c c c c c c}
  \hline\noalign{\smallskip}
Emission region &  EW(H$\alpha$),  & $\lg${[NII]$\lambda$6583/H$\alpha$} &   12+$\lg$(O/H)           &  12+$\lg$(O/H)              & $\Sigma$(SFR(H$\alpha$)),         \\
                &       \AA\       &                                      & {\it N2} indicator         & {\it N2} indicator           & $M_{\odot}$~yr$^{-1}$~kpc$^{-2}$  \\
                &                  &                                   & (\citealt{Pettini+Pagel+2004}) & (\citealt{Marino+etal+2013}) &                                   \\
\hline
PGC 48114       &                  &                                      &                            &                              &                                   \\ 
O               &  $0.23 \pm 0.05$ &           $0.40 \pm 0.11$            &                            &                              &                                   \\
1               & $15.33 \pm 0.26$ &          $-0.52 \pm 0.07$            &     $8.60 \pm 0.04$        &  $8.50 \pm 0.04$             & $(1.9 \pm 0.6) \times 10^{-3}$    \\
2               & $10.43 \pm 0.80$ &          $-0.29 \pm 0.04$            &                            &                              &                                   \\
3               & $17.11 \pm 0.46$ &          $-0.35 \pm 0.04$            &     $8.70 \pm 0.02$        &  $8.58 \pm 0.02$             & $(2.1 \pm 0.6) \times 10^{-3}$    \\
4               & $10.60 \pm 0.63$ &          $-0.32 \pm 0.04$            &                            &                              &                                   \\
5               &  $4.40 \pm 0.30$ &          $-0.37 \pm 0.04$            &     $8.69 \pm 0.02$        &  $8.57 \pm 0.02$             & $(1.5 \pm 0.6) \times 10^{-3}$    \\
6               &  $8.00 \pm 0.14$ &          $-0.41 \pm 0.04$            &     $8.67 \pm 0.02$        &  $8.55 \pm 0.02$             & $(1.2 \pm 0.6) \times 10^{-3}$    \\
7               &  $9.34 \pm 0.28$ &          $-0.40 \pm 0.05$            &     $8.67 \pm 0.03$        &  $8.56 \pm 0.03$             & $(1.0 \pm 0.5) \times 10^{-3}$    \\
8               &  $2.69 \pm 0.21$ &          $-0.27 \pm 0.05$            &                            &                              &                                   \\
\hline
Entire disk of  &                  &                                      &                            &                              &                                   \\
PGC~48090       & $13.68 \pm 0.16$ &          $-0.91 \pm 0.09$            &     $8.38 \pm 0.05$        &  $8.32 \pm 0.04$             & $(2.11 \pm 0.05) \times 10^{-3}$  \\
\noalign{\smallskip}\hline
\end{tabular}
\ec
\end{table}

As can be seen from Table~\ref{tab5}, the metallicity of the gas in star-forming clumps of the S0 galaxy PGC~48114 has a 
near-solar value, while the average oxygen abundance in the disk of the late-type galaxy PGC~48090 is 0.3\,dex less than 
the solar value, which is consistent with data for other galaxies of this type (\citealt{Pettini+Pagel+2004}).
The sizes of the selected clumps, corrected for the seeing, varies in the range from 1.7 arcsec to 2.6 arcsec, 
corresponding to the physical sizes of 790 -- 1200 pc (at the adopted distance to the galaxy of 96.6~Mpc). 
It is also worth noting that in PGC~48114, there is no current SF in the central clump O, that is consistent with our 
spectral data. Since most of the star-forming clumps are located in the northwestern arc of the ring of PGC~48114, this local 
SF was most likely induced by the close flyby of PGC~48090 from northeast to southwest, and now located west of PGC~48114.

\subsection{The structure of ESO~234-011 and ESO~444-076}
\label{structure}

By inspecting the deep images of ESO~234-011 and ESO~444-076 presented by the Legacy Survey service (\citealt{Dey+etal+2019}),
one can assure that the blue rings are embedded into the more extended reddish stellar disks (Fig.~\ref{fig1}). We have undertaken 
isophotal analysis of these deep images in the $r$-band and then have calculated azimuthally-averaged surface brightness profiles. 
The outer surface brightness profiles were averaged over the elliptic apertures with the shape and orientation taken from the 
outermost estimates of the ellipticity and major-axis position angles, so assuming that the outer disks are circular.

\begin{figure}
   \centering
  \includegraphics[width=\textwidth]{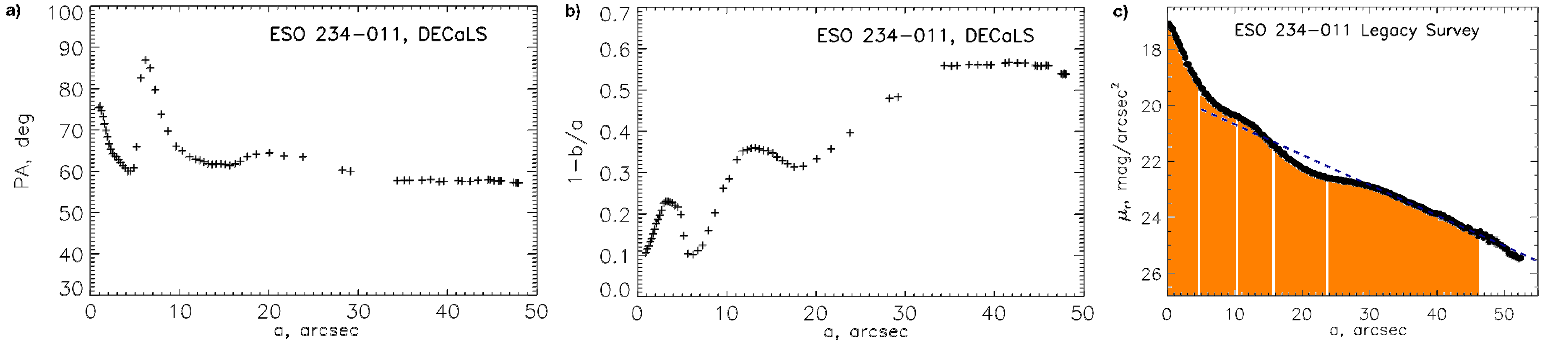}
   \caption{The results of isophote analysis and the surface brightness profile in the $r$-band azimuthally-averaged 
in elliptical rings for ESO~234-011; for the outer disk, the fitted exponential law is also shown. 
The ranges over which the Lick indices were averaged are shown in orange.}
   \label{fig10}
   \end{figure}

Figure~\ref{fig10} presents the results of isophotal analysis and the surface brightness radial profile for ESO~234-011. 
Though the galaxy is classified as an unbarred one, we see a local turn of the position angle at a radius of 
$R = 6$ arcsec -- just where the shock-induced emission regions are observed (Fig.~\ref{fig6}). The ellipticity
profile reveals two local maxima, at $R\approx 4$~arcsec and $R\approx 13$~arcsec betraying a diversity
of the inner non-axisymmetric structures. The ellipticity plateau, and so an exponential stellar disk, starts only at
$R\approx 33$~arcsec, or at 12~kpc from the center.

\begin{figure}
   \centering
  \includegraphics[width=\textwidth]{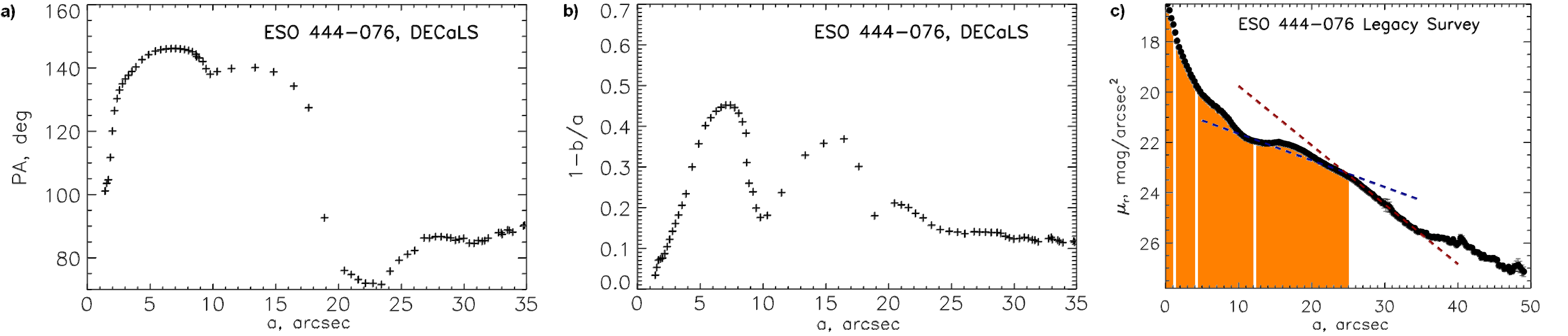}
   \caption{The results of isophote analysis and the surface brightness profile in the $r$-band azimuthally-averaged 
in elliptical rings for ESO~444-076; for the inner and outer disks, the fitted exponential laws are also shown. 
The ranges over which the Lick indices were averaged are shown in orange.}
   \label{fig11}
   \end{figure}

Figure~\ref{fig11} presents the results of isophotal analysis and the surface brightness radial profile for ESO~444-076. 
The results of the isophotal analysis reveal two local maxima of the ellipticity at the bar edge and at the ring radius, 
both structures are turned by some 70~deg with respect to the line of nodes. The surface brighness profile (Fig.~\ref{fig11}c) 
are of II.o$+$III type (\citealt{Gutierrez+etal+2011}), and the ring is embedded into the innermost segment of the stellar disk.

A similar isophote analysis and the surface brightness profile in the $r$-band for the galaxy PGC~48114 have been 
performed and presented by us previously (see fig.~2 in \cite{Proshina+etal+2019}).

\subsection{Properties of stellar populations of S0 galaxies in isolated pairs}

\begin{figure}
   \centering
  \includegraphics[width=\textwidth]{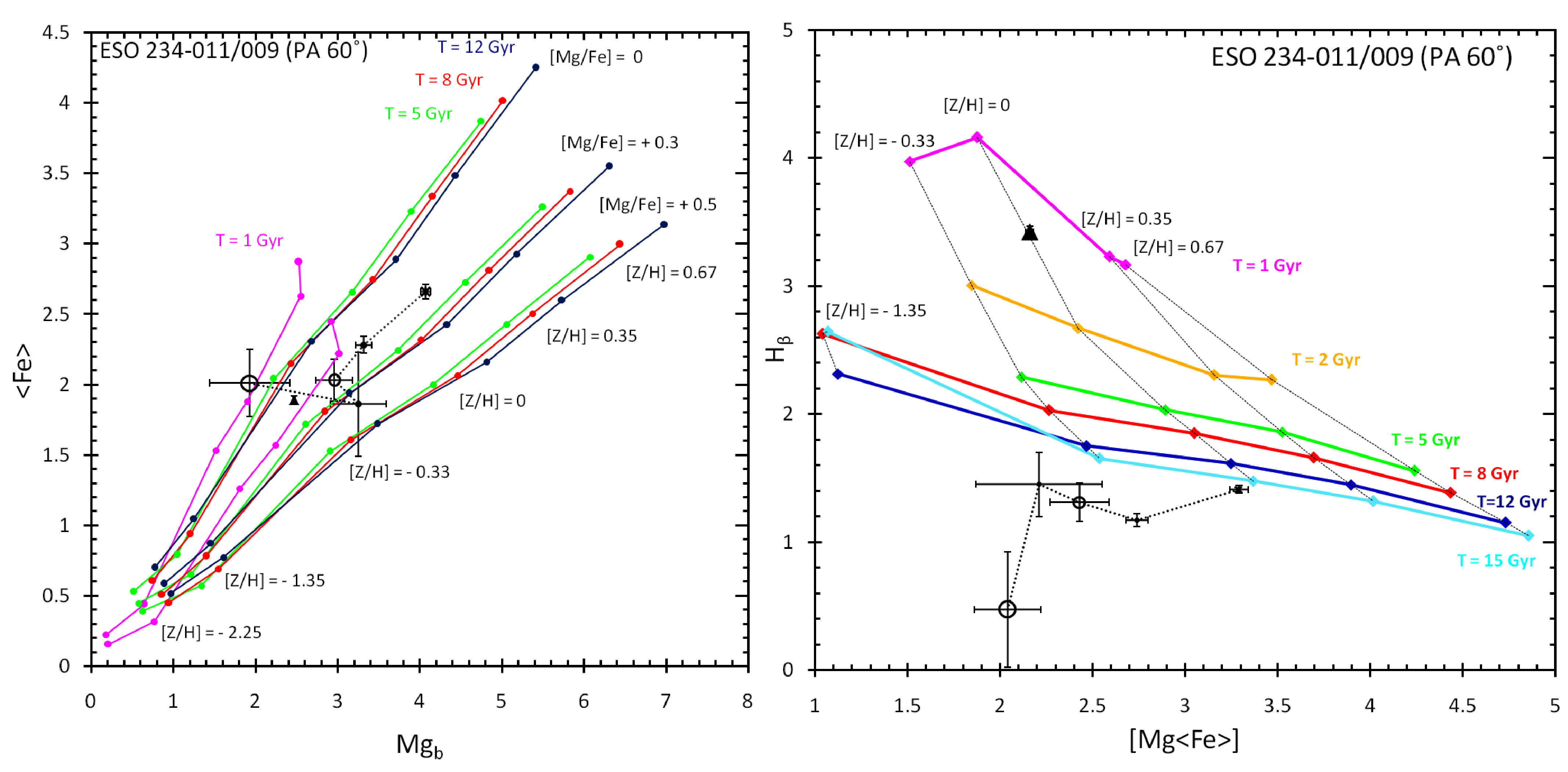}
   \caption{Diagnostic Lick index-index diagrams for determining the properties of stellar populations in galaxies ESO~234-011 ({\it dots} 
and {\it open circles}) and ESO~234-009 ({\it triangle}).
The colored lines indicate the sequences of equal ages, the gray lines -- the sequences of equal metallicity.}
   \label{fig12}
   \end{figure}

\begin{figure}
   \centering
  \includegraphics[width=\textwidth]{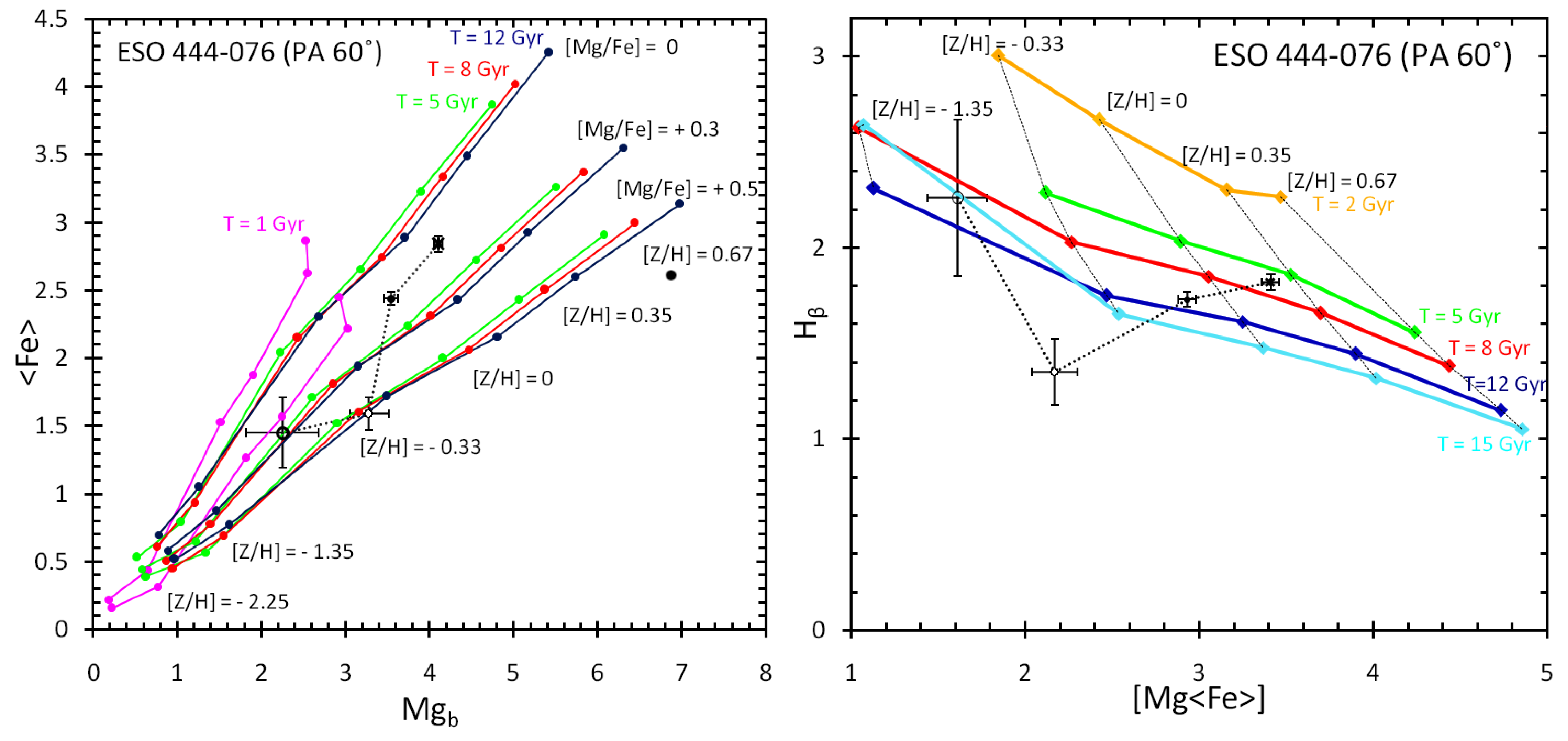}
   \caption{Diagnostic Lick index-index diagrams for determining the properties of stellar populations in the galaxy ESO~444-076.
The colored lines indicate the sequences of equal ages, the gray lines -- the sequences of equal metallicity.}
   \label{fig13}
   \end{figure}

As a tool to analyze the properties of the stellar population of the galaxies, we have used the measurements of a set of Lick indices: 
H$\beta$, Mg$_b$, Fe5270, and Fe5335. The Lick indices characterizing absorption lines in the integrated stellar population spectrum 
represent equivalent widths calculated under specific conditions. Namely, the continuum above every line is estimated by linear 
interpolation between ‘blue side’ and ‘red side’ wavelength ranges whose borders are fixed strictly in the restframe. The full 
description of the Lick index system including about twenty the strongest optical absorption lines was formulated by Guy Worthey with 
co-authors in their famous paper, \cite{Worthey+etal+1994}. There a full table with the line borders fixed and continuum range borders 
fixed can be found. We have calculated the instrumental Lick indices H$\beta$, Mg$_b$, Fe5270, and Fe5335 for our galaxies along the 
slit by using these prescriptions. Earlier a set of standard stars from \cite{Worthey+etal+1994} were observed by us with the same RSS 
instrumental setup, and the original instrumental Lick indices were calculated also for the standard stars. This allowed us 
(\citealt{Katkov+etal+2015}) to establish relationship between instrumental SALT/RSS index system and the standard Lick one.
So we converted our derived instrumental SALT/RSS Lick indices to the standard Lick system (\citealt{Worthey+etal+1994}) using the 
conversion formulas for the RSS spectrograph of the SALT telescope from \cite{Katkov+etal+2015}. We made corrections for the H$\beta$ 
emission using fluxes in H$\alpha$ and assuming the Balmer decrement H$\alpha$/H$\beta$ = 4, i.e., $\Delta$H$\beta$ = 0.25EW(H$\alpha$) 
(\citealt{Stasinska+Sodre+2001}), since the ionized gas in the galaxies ESO~234-011 and ESO~444-076 along the long slit is shock-excited. 
For ESO~234-009, the Balmer decrement was calculated from the measured fluxes of the H$\alpha$ and H$\beta$ emission lines.
After that, we compared the derived Lick indices with the Simple Stellar Population models grids of \cite{Thomas+etal+2003}.
To trace how the average age and metallicity of the stellar population change with distance from the center (along the 
long slit), we averaged the values of the Lick indices over radial intervals in which their values remain approximately constant 
or change smoothly, and boundaries of these intervals often coincide with the structural features of the galaxies, which we also 
took into account. Thus, for the galaxy ESO~234-011 the following intervals, shaded in orange in Figure~\ref{fig10}c, 
were chosen: 0 -- 4.5 arcsec (bulge), 5 -- 10 arcsec (lens), 10.5 -- 15.5 arcsec (inner ring), 16 -- 23.5 arcsec, 24 -- 46 arcsec (outer 
ring and disk). For the galaxy ESO~444-076, the Lick indices were averaged over the intervals, shaded in orange in 
Figure~\ref{fig11}c: 0 -- 1 arcsec (nucleus), 1.5 -- 4 arcsec (bulge), 4.5 -- 12 arcsec (inner ring), and 12.5 -- 25 arcsec (outer ring).
The diagnostic diagrams $<$Fe$>$ vs. Mg$_b$ and H$\beta$ vs. [Mg$_b$Fe] for the galaxies ESO~234-011 and ESO~444-076 are shown in 
Figure~\ref{fig12} and Figure~\ref{fig13}, respectively. 
In the plots, the rings are indicated by open circles of different diameters (the inner one is small, the outer one is larger), 
the center is a bold symbol, and the remaining values along the slit are indicated by black dots connected by a dotted line; 
the values for the companion galaxy ESO~234-009 are indicated by a triangle (with error bars within this symbol).

Analyzing the diagnostic diagrams above, we can conclude that galaxy ESO~234-011 has a main stellar population 
typical for early-type galaxies: the average age of stars is approximately 15 Gyr, the metallicity of stars at 
the center reaches solar value, and decreases with distance from the center. The [Mg/Fe] ratio of $+$0.2\,dex, constant 
within 15 arcsec from the center, indicates a relatively short star formation burst that occurred during the galaxy's 
formation approximately 15 Gyr ago. The observed increase in [Mg/Fe] after 15 arcsec to $+$0.4\,dex, on the one hand, 
is consistent within the error bars with the value of $+$0.2\,dex, on the other hand, may indicate a possible brief burst of 
star formation that occurred some time ago and led to the formation of a small amount of young stellar population 
that did not significantly contribute to the predominantly old stellar population of the galaxy. In the outer disk region 
(24 -- 46 arcsec), where the ring is currently visible only in optical wavelengths, the [Mg/Fe] ratio drops below solar value, 
the stellar metallicity remains significantly below solar value, and the stellar population is just as old.
Such underabundance of oxygen relative to iron due to small number of bursts of SF, separated by extended 
quiescent phases, are typical for dwarf galaxies, e.g., LMC (\citealt{Gilmore+Wyse+1991}). This point outs that outer ring 
and disk of ESO~234-011 consist of material from merged dwarf satellite, or on the intermittent nature of SF 
in the ring caused by intermittent accretion of gas from the outside. 
However, the stellar population in the neighboring late-type galaxy ESO~234-009 (indicated by the {\it triangle} in 
Fig.~\ref{fig12}) has an average age of less than 2~Gyr, a solar metallicity, and a [Mg/Fe] ratio of $+$0.2\,dex.

In the S0 galaxy ES0~444-076, the stellar population in the circumnuclear region has an average age of approximately 6~Gyr, 
a [Mg/Fe] ratio of $+$0.15\,dex, and a metallicity approximately twice the solar value, indicating a recent rejuvenation episode. 
However, at a distance of 4 arcsec from the galaxy's center, the average age of stars increases to 10~Gyr, the metallicity 
drops below the solar value, and the [Mg/Fe] ratio remains at $+$0.15\,dex. At distances of 4.5 to 12 arcsec from the center, 
the [Mg/Fe] ratio increases to $+$0.5\,dex, and then decreases to $+$0.3\,dex in the outer ring region, with the metallicity of 
stars several times lower than the solar value.

The same diagnostic diagrams for the galaxy PGC~48114 are given in our previous paper (\citealt{Proshina+etal+2019}, fig.~9). 
In the center of the S0 galaxy PGC~48114, the stellar population has an average age of about 5~Gyr, a [Mg/Fe] ratio of 
about $+$0.1\,dex, and a metallicity approximately twice the solar value, indicating a recent rejuvenation episode. 
As the distance from the center increases, the metallicity and [Mg/Fe] decrease and the age of the stellar population 
increases. In the outer ring of the galaxy PGC~48114, the [Mg/Fe] ratio drops below solar value and the stellar metallicity 
significantly below solar value. This again indicates rare episodes of SF separated by extended quiescent phases 
(\citealt{Gilmore+Wyse+1991}).

Thus, it is clear that in all three S0 galaxies the stellar population in the outer rings differs significantly in its 
properties from the main stellar population of S0 galaxies and was probably formed as a result of accretion from outside 
stars and/or gas.

\subsection{Estimation of recent star formation rates}
\label{SFR}

\begin{figure}
   \centering
  \includegraphics[width=\textwidth]{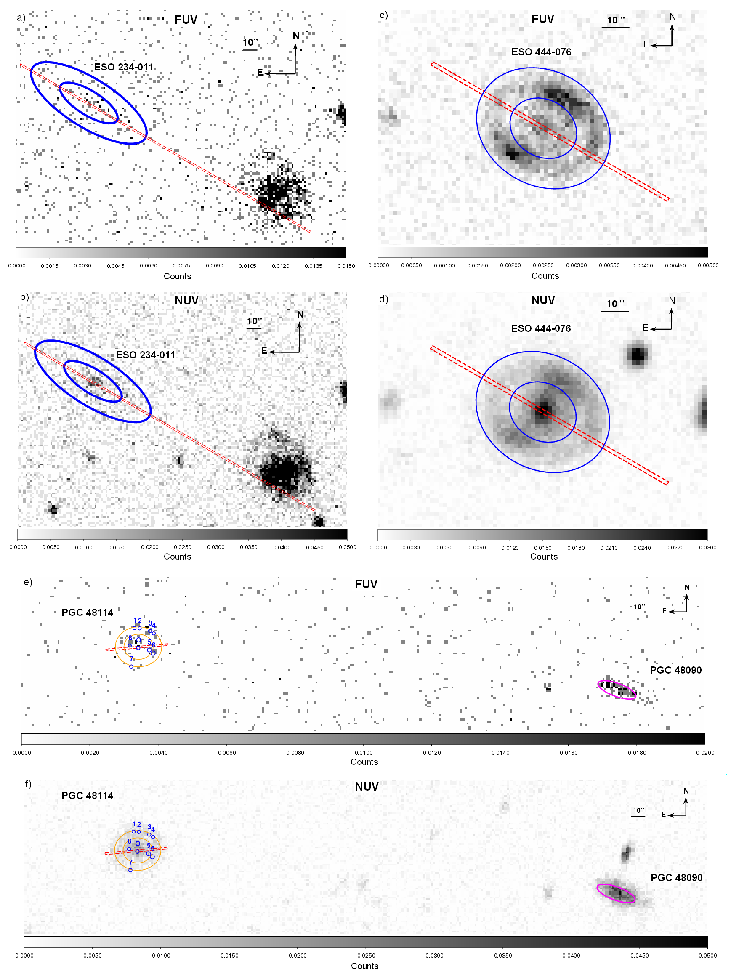}
   \caption{FUV and NUV images with sites of annular apertures and long slits superimposed for 
ESO~234-011 ({\it a, b}), ESO~444-076 ({\it c, d}), and the pair of galaxies PGC~48114 and PGC~48090 ({\it e, f}).}
   \label{fig14}
   \end{figure}

As seen in the optical images (Fig.~\ref{fig1}), all three S0 galaxies have outer rings. In the UV images of these galaxies 
(Fig.~\ref{fig14}) obtained with the {\it GALEX} space telescope, a very bright UV ring, coinciding with the optical ring, 
is visible in the galaxy ESO~444-076; the rings of the other two S0 galaxies are barely visible in the UV.

Fluxes in the FUV and NUV bands are indicators of star formation on timescales over 100 and 200 Myr ago, respectively. 
We measured them in apertures corresponding to the optical rings using several {\it GALEX} UV images listed in Table~\ref{tab6} 
and retrieved from the Mikulski Archive for Space Telescopes\footnote{\url{https://galex.stsci.edu/GR6}}.

\begin{table}
\bc
\caption[]{The log of UV observations of S0 galaxies by {\it GALEX}\label{tab6}}
\setlength{\tabcolsep}{4pt}
\small
 \begin{tabular}{l c c c c}
  \hline\noalign{\smallskip}
Galaxy        & Band & Program identifier    & Date       & Exposure, s \\
\hline
PGC~48114     & FUV  & AIS~221~0001~sg12     & 2004-04-16 &   112       \\ 
              & NUV  & AIS~221~0001~sg12     & 2004-04-16 &   112       \\
              & NUV  & MISGCSN3~33572~0222   & 2011-05-05 &   1684.1    \\
\hline
ESO~234-011   & FUV  & AIS~367~0001~sg89     & 2003-08-05 &   145       \\
              & NUV  & AIS~367~0001~sg89     & 2003-08-05 &   145       \\
\hline
ESO~444-076   & FUV  & AIS~351~0001~sg68     & 2007-04-20 &   197       \\
              & FUV  & NGA~M83~582~F5~19~158 & 2003-06-07 &   1340.05   \\
              & FUV  & GI3~050007~NGC5236    & 2007-03-15 &   14250.4   \\
              & NUV  & AIS~351~0001~sg68     & 2007-04-20 &   197       \\
              & NUV  & NGA~M83~582~F5~19~158 & 2003-06-07 &   1349.05   \\
              & NUV  & NGA~M83~west          & 2006-04-25 &   1406      \\
              & NUV  & GI3~050007~NGC5236    & 2007-03-15 &   14256.4   \\
\noalign{\smallskip}\hline
\end{tabular}
\ec
\end{table}

To account for dust absorption in the galaxies under study when calculate star formation rates (SFRs) using formulas of 
\cite{Kennicutt+Evans+2012}, we measured the fluxes in the IR band (22 $\mu$m) using IR images obtained by the {\it WISE} 
space telescope in the W4 band and taken from the NASA/IPAC Infrared Science Archive 
(IRSA\footnote{\url{https://irsa.ipac.caltech.edu/applications/wise}}).
We also corrected for extinction in our Galaxy in the direction of the galaxies under study, taking $A_B = 0.091$\,mag, 
$A_B = 0.170$\,mag, and $A_B = 0.136$\,mag, for PGC~48114, ESO~444-076, and ESO~234-011, respectively (according to 
data from \cite{Schlafly+Finkbeiner+2011} from NED).
The calculated integral SFRs in the ring structures of S0 galaxies, based on measured fluxes in the FUV 
and NUV bands, as well as the surface densities of these SFRs, are given in Table~\ref{tab7}.

\begin{table}
\bc
\caption[]{Mean surface densities of SFRs and integral SFRs in the rings of S0 galaxies\label{tab7}}
\setlength{\tabcolsep}{4pt}
\small
 \begin{tabular}{l c c c c c}
  \hline\noalign{\smallskip}
Galaxy      & Ring boundaries &        $\Sigma_{SFR(FUV)}$        &         $\Sigma_{SFR(NUV)}$       &    SFR(FUV)           &     SFR(NUV)          \\
            &     (arcsec)    & $M_{\odot}$ yr$^{-1}$ kpc$^{-2}$  & $M_{\odot}$ yr$^{-1}$ kpc$^{-2}$  & $M_{\odot}$ yr$^{-1}$ & $M_{\odot}$ yr$^{-1}$ \\
\hline
ESO~234-011 &    24 -- 48     & $(1.54 \pm 0.16) \times 10^{-4}$  & $(1.62 \pm 0.08) \times 10^{-4}$  & $0.113 \pm 0.012$ & $0.119 \pm 0.006$         \\ 
ESO~444-076 &   12.5 -- 25    & $(6.68 \pm 0.11) \times 10^{-4}$  & $(8.22 \pm 0.09) \times 10^{-4}$  & $0.074 \pm 0.001$ & $0.091 \pm 0.001$         \\
PGC~48114   &    10 -- 16     & $(4.43 \pm 0.37) \times 10^{-4}$  & $(6.73 \pm 0.52) \times 10^{-4}$  & $0.048 \pm 0.004$ & $0.073 \pm 0.006$         \\
\noalign{\smallskip}\hline
\end{tabular}
\ec
\end{table}

\begin{figure}
   \centering
  \includegraphics[width=10cm, angle=0]{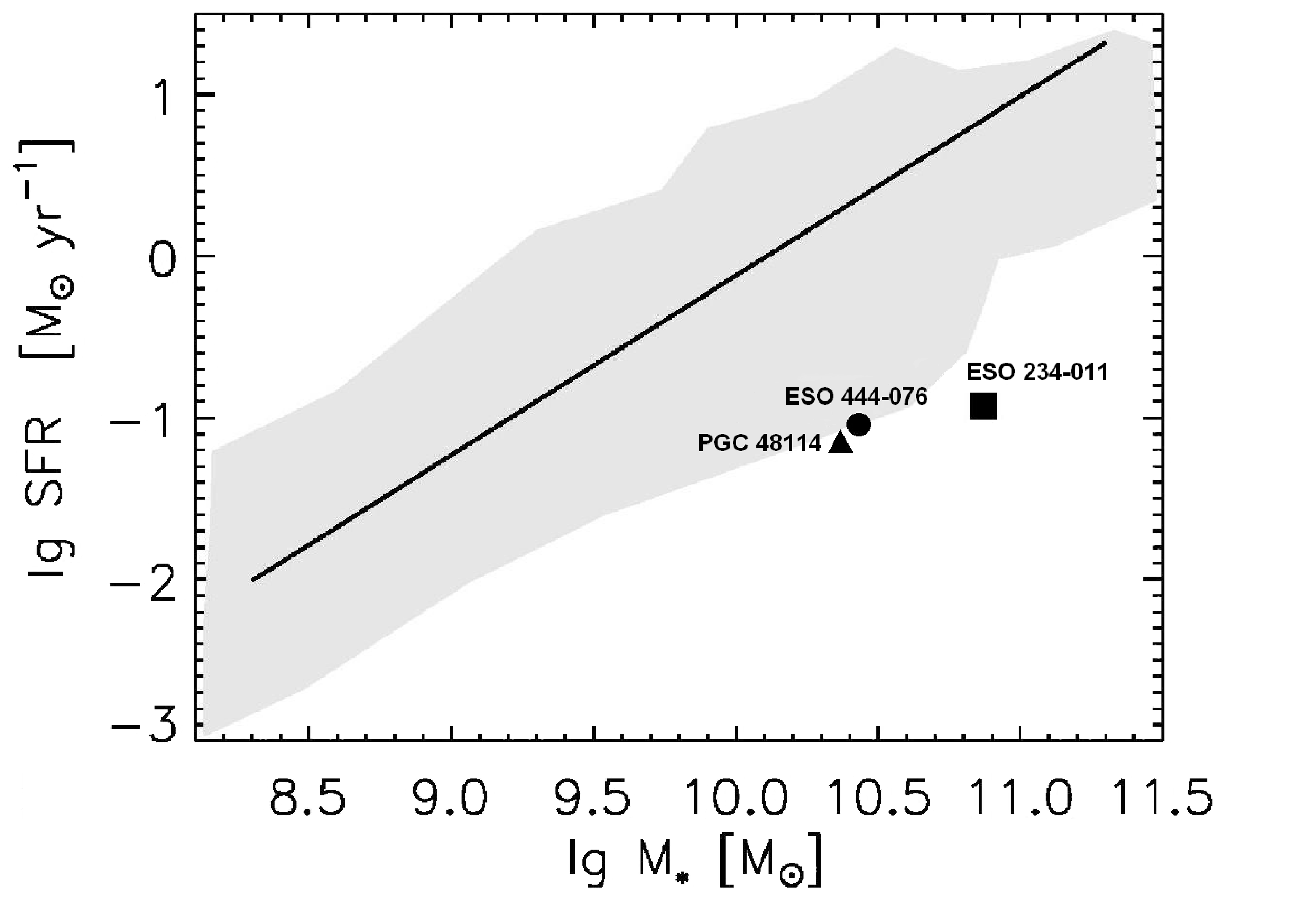}
   \caption{The position of three S0 galaxies on the scaling relation from \cite{Naluminsa+etal+2021} based on data from the WHISP survey. 
All three S0 galaxies are well below the main sequence denoted by {\it solid line}. The gray shadow regions mark the scatter of data from 
the original paper.}
   \label{fig15}
   \end{figure}

Among the three S0 galaxies, the galaxy ESO~234-011 has the highest integral SFR over the past 200 Myr -- 
slightly more than 0.1~$M_{\odot}$~yr$^{-1}$, but the surface densities of SFRs are the lowest among the three 
galaxies. Since the SFRs calculated from the FUV and NUV fluxes are practically equal and there is no current 
SF (according to our long-slit spectroscopy data), it can be concluded that SF in ESO~234-011 occurred 
about 200~Myr ago, was fleeting, and is ending now. This conclusion generally applies to the other two S0-type galaxies. 
Moreover, in PGC~48114, we found local clumps of current SF, mostly in the northwestern arc.
And in ESO~444-076, the surface densities of SFRs are the highest among the three S0 galaxies, which is why it has 
such a bright ring in the UV images with characteristic UV brightenings (ansae) near the ends of the bar.
However, when regarding the galaxies with respect to the main sequence (Fig.~\ref{fig15}), with their SFR estimated by us 
over the last 200~Myr, we must conclude that these three S0 galaxies are globally quenched.

For comparison, Table~\ref{tab8} presents the mean (averaged over the disk area) surface densities of SFRs and the 
integral SFRs in late-type companion galaxies. As can be seen, the mean surface densities of SFRs over 
a disk in these galaxies are a several times higher than in the rings of nearby S0 galaxies. But the surface densities of the current 
rates of star formation in the star-forming clumps in PGC~48114 and the average for the disk of PGC~48090 (Table~\ref{tab5}, last column) 
are similar in values.

\begin{table}
\bc
\caption[]{Mean surface densities of SFRs and integral SFRs in the disks of LTGs\label{tab8}}
\setlength{\tabcolsep}{4pt}
\small
 \begin{tabular}{l c c c c c}
  \hline\noalign{\smallskip}
Galaxy      & $r_a \times r_b$ of disk &        $\Sigma_{SFR(FUV)}$         &         $\Sigma_{SFR(NUV)}$        &       SFR(FUV)        &     SFR(NUV)          \\
            &       (arcsec)           & $M_{\odot}$ yr$^{-1}$ kpc$^{-2}$   & $M_{\odot}$ yr$^{-1}$ kpc$^{-2}$   & $M_{\odot}$ yr$^{-1}$ & $M_{\odot}$ yr$^{-1}$ \\
\hline
ESO~234-009 & $29.5 \times 25.1$       & $(3.208 \pm 0.080) \times 10^{-3}$ & $(3.491 \pm 0.080) \times 10^{-3}$ &   $1.067 \pm 0.020$   & $1.161 \pm 0.030$     \\ 
ESO~444-077 & $49.5 \times 38.3$       & $(0.540 \pm 0.008) \times 10^{-3}$ & $(0.667 \pm 0.006) \times 10^{-3}$ &   $0.198 \pm 0.003$   & $0.244 \pm 0.002$     \\
PGC~48090   & $14.0 \times 4.6$        & $(2.315 \pm 0.223) \times 10^{-3}$ & $(2.447 \pm 0.195) \times 10^{-3}$ &   $0.098 \pm 0.013$   & $0.105 \pm 0.007$     \\
\noalign{\smallskip}\hline
\end{tabular}
\ec
\end{table}

The current rate of star formation, calculated from the integral H$\alpha$ emission line flux from the entire disk 
of PGC~48090, is $0.088 \pm 0.002$~$M_{\odot}$~yr$^{-1}$. And the mean surface density of the current SFR, 
estimated from the flux in H$\alpha$, is $0.00211 \pm 0.00005$~$M_{\odot}$~yr$^{-1}$~kpc$^{-2}$. These values are 
slightly less than those calculated on the timescales over 100 and 200 Myr ago from the integral fluxes in FUV and NUV 
bands, respectively (Table~\ref{tab8}). This indicates that the modern episode of SF is lasting hundreds Myr, 
and it may have been forced by a close flyby near its neighbor galaxy PGC~48114 a several hundred Myr ago. In general, such 
characteristics of SF are typical of normal disk and irregular galaxies (see fig.~9 in \citealt{Kennicutt+Evans+2012}).

\section{Discussion about the origin of gas}
\label{Discussion} 

Our observational data indicate the presence of ionized gas in the three lenticular galaxies under study and 
kinematic hints of its external origin in two of them. However, our data do not allow us to unambiguously 
identify a specific gas source in each case. Only in the galaxy ESO~234-011 does the observed jump in the line-of-sight 
velocities from the side of the neighboring galaxy ESO~234-009, traced by the H$\alpha$ emission line, indicate that 
the neighboring galaxy ESO~234-009 is the most likely source of the gas. As shown in Table~\ref{tab2}, all companion 
late-type galaxies contain significant amounts of atomic hydrogen. However, as mentioned in a recent paper by 
\cite{Sil'chenko+etal+2023}, in addition to gas inflow from neighboring galaxies or their complete merger, an inflow 
of cold gas from cosmological filaments (\citealt{Dekel+Birnboim+2006, Dekel+etal+2009}), or the settling of 
cooling gas from hot galactic halos (\citealt{Keres+etal+2005, Joung+etal+2012}), are possible, 
or several galaxies may share a common extended reservoir of atomic hydrogen (\citealt{Sancisi+etal+2008}).

According to ATCA observations by \cite{Oosterloo+etal+2007}, the atomic hydrogen mass in the galaxy ESO~234-011 is 
less than $1.1 \times 10^9$~$M_{\odot}$. However, the authors do not provide an \ion{H}{I} distribution map for this galaxy. 
Such a map, as well as line-of-sight velocity and velocity dispersion maps covering both galaxies in a pair and 
their surroundings, would help clarify the source of the gas in both galaxies in each pair under consideration.

Short-lived (on galactic scales) and/or local bursts of star formation are likely caused by intermittent gas flows, 
for example, during the approach of neighboring galaxies that form isolated pairs, or as a result of rare minor mergers. 
Also the accreting gas requires time to lose angular momentum and settle in the plane of the galactic disk, after which 
SF begins. After the settled gas is exhausted, SF ceases until the gas again settles in the plane 
of the disk in the proper quantity.

\section{Conclusions}
\label{Conclusions} 

For three lenticular galaxies (ESO~234-011, ESO~444-076, PGC~48114) that are part of isolated pairs, where the 
companions are late-type galaxies containing the proper amount of atomic hydrogen, we carried out long-slit spectroscopy, 
and for one pair of the three, we also performed narrow-band photometry in the emission lines H$\alpha$ and 
[NII]$\lambda$6583. We analyzed the obtained data and also involved publicly available archival optical, UV, and IR 
images, and came to the following conclusions.

\begin{itemize}
\item The spectra of three lenticular galaxies contain emission lines of ionized gas, which means that gas is present there.

\item In the galaxy PGC~48114, the mechanism for excitation of ionized gas varies in different parts of the ring 
(although the kinematics of ionized gas and stars are generally consistent). Narrow-band images in the H$\alpha$ and 
[NII]$\lambda$6583 emission lines, obtained for this galaxy and its companion using the MaNGaL mapper on the 2.5-m 
telescope of the Caucasus Mountain Observatory (CMO) of the SAI MSU, revealed an arc of the sites of current star formation 
in the northwestern part of the ring, apparently induced by a recent flyby of a neighboring galaxy located to the west and 
slightly south, in which we detected the brightest H$\alpha$ clump on the northern side of the disk.

\item In galaxies ESO~234-011 and ESO~444-076, ionized gas along the long slit is shock-excited and have kinematic discrepancy with 
the stars, which is may be evidence of matter accretion at an angle to the main disk of S0 galaxies and, consequently, may inhibits 
ongoing star formation in them (as in some other galaxies with misaligned kinematics discussed in \citealt{Sil'chenko+etal+2019}).

\end{itemize}

To clarify the origin of the gas in the S0 galaxies of these isolated pairs, maps of the distribution of atomic hydrogen, 
as well as maps of line-of-sight velocities and velocity dispersions, are needed.

\normalem
\begin{acknowledgements}

The study was supported by the grant from the Russian Science Foundation No.~22-12-00080-$\Pi$. 
A.Yu.K. acknowledges support from the National Research Foundation (NRF) of South Africa.

The authors thank A.V. Moiseev for the primary processing of narrowband observational data obtained by MaNGaL. 

This work is based on observations conducted with the 10-m Southern African Large Telescope (SALT) of the 
South African Astronomical Observatory, programs 2012-2-RSA\_OTH-002 and 2018-1-MLT-003 (PI: Kniazev), 
as well as the 2.5-m telescope of the Caucasus Mountain Observatory (CMO) of the Sternberg Astronomical Institute of the 
Moscow State University, program GAL\_PROSH (PI: Proshina). 
We used the public data of the Legacy Survey (http://legacysurvey.org) and {\it GALEX} (http://galex.stsci.edu/GR6) survey. 
The Legacy Survey imaging of the DESI footprint is supported by the Director, Office of Science, Office of High Energy 
Physics of the U.S. Department of Energy under Contract No. DE-AC02-05CH1123; by the National Energy Research Scientific 
Computing Center, a DOE Office of Science User Facility under the same contract; and by the U.S. National Science Foundation, 
Division of Astronomical Sciences under Contract No. AST-0950945 to NOAO.
The NASA {\it GALEX} mission data were retrieved from the Mikulski Archive for Space Telescopes (MAST) at the Space Telescope 
Science Institute, which is operated by the Association of Universities for Research in Astronomy, Inc., under NASA 
contract NAS 5-26555. The {\it WISE} image used by us was taken from the NASA/IPAC Infrared Science Archive, which is operated 
by the California Institute of Technology, under contract with the National Aeronautics and Space Administration. 
We also used the NASA/IPAC Extragalactic Database (NED) which is operated by the California Institute of Technology, 
under contract with the National Aeronautics and Space Administration. 
We acknowledge the usage of the Lyon Extragalactic Database (HyperLEDA, http://atlas.obs-hp.fr/hyperleda/).

\end{acknowledgements}

\bibliographystyle{raa}
\bibliography{Iso_pairs}

\end{document}